\documentclass[5p,twocolumn]{elsarticle}

\usepackage{amssymb}
\usepackage{latexsym}
\usepackage{framed,multirow}

\usepackage{url}
\usepackage{float}
\usepackage{placeins}
\usepackage{xcolor}
\usepackage{makecell}
\usepackage{tabu, booktabs}
\usepackage{lineno}
\usepackage{bigstrut}
\usepackage{enumitem}
\definecolor{newcolor}{rgb}{.8,.349,.1}
\usepackage{colortbl}
\usepackage[table]{xcolor}
\usepackage{subcaption}

\begin{document}

\begin{frontmatter}

\title{\textit{SegRap2025}: A Benchmark of Gross Tumor Volume and Lymph Node Clinical Target Volume \textit{Seg}mentation for \textit{Ra}diotherapy \textit{P}lanning of Nasopharyngeal Carcinoma}%

\author[1,2,3]{Jia Fu}
\author[1]{Litingyu Wang}
\author[1]{He Li}
\author[1]{Zihao Luo}
\author[1]{Huamin Wang}
\author[4]{Chenyuan Bian}
\author[5]{Zijun Gao}
\author[5]{Chunbin Gu}
\author[6]{Xin Weng}
\author[7]{Jianghao Wu}
\author[8]{Yicheng Wu}
\author[7]{Jin Ye}
\author[9]{Linhao Li}
\author[9]{Yiwen Ye}
\author[9]{Yong Xia}
\author[10]{Elias Tappeiner}
\author[11]{Fei He}
\author[12]{Abdul qayyum}
\author[13]{Moona Mazher}
\author[12]{Steven A Niederer}
\author[14]{Junqiang Chen}
\author[15]{Chuanyi Huang}
\author[15]{Lisheng Wang}
\author[16]{Zhaohu Xing}
\author[16]{Hongqiu Wang}
\author[16]{Lei Zhu}
\author[2]{Shichuan Zhang}
\author[1,3]{Shaoting Zhang}
\author[2]{Wenjun Liao\corref{cores}}\ead{lwjpsy@163.com}
\author[1,3]{Guotai Wang\corref{cores}}\ead{guotai.wang@uestc.edu.cn}
\cortext[cores]{Corresponding authors.}

\address[1]{School of Mechanical and Electrical Engineering, University of Electronic Science and Technology of China, Chengdu, China.}
\address[2]{Department of Radiation Oncology, Sichuan Cancer Center, Radiation Oncology Key Laboratory of Sichuan Province, Sichuan Clinical Research Center for Cancer, Sichuan Cancer Hospital and Institute, University of Electronic Science and Technology of China, Chengdu, China.}
\address[3]{Shanghai AI Lab, Shanghai, China.}
\address[4]{The Affiliated Hospital of Qingdao University, Qingdao, China.}
\address[5]{Department of Computer Science and Engineering, The Chinese University of Hong Kong, Hong Kong, China.}
\address[6]{Bank of China, Cangshan Sub-branch, Fujian Branch, Fuzhou, China.}
\address[7]{Faculty of Information Technology, Monash University, Melbourne, Australia.}
\address[8]{Department of Computing and Department of Brain Sciences, Imperial College London, United Kingdom.}
\address[9]{School of Computer Science and Engineering, Northwestern Polytechnical University, Xi'an, China.}
\address[10]{UMIT Tirol - Private University for Health Sciences and Health Technology, Austria.}
\address[11]{School of Information and Communication Engineering, University of Electronic Science and Technology of China, Chengdu, China.}
\address[12]{National Heart and Lung Institute, Imperial College London, London, United Kingdom.}
\address[13]{Hawkes Institute, Department of Computer Science, University College London, London, United Kingdom.}
\address[14]{Shanghai MediWorks Precision Instruments Co., Ltd., China.}
\address[15]{School of Automation and Intelligent Sensing, Shanghai Jiao Tong University, Shanghai, China.}
\address[16]{Department of Systems Hub, Hong Kong University of Science and Technology (Guangzhou), Guangzhou, China.}

\begin{abstract}
Accurate delineation of Gross Tumor Volume (GTV), Lymph Node Clinical Target Volume (LN CTV), and Organ-at-Risk (OAR) from Computed Tomography (CT) scans is essential for precise radiotherapy planning in Nasopharyngeal Carcinoma (NPC). Building upon SegRap2023, which focused on OAR and GTV segmentation using single-center paired non-contrast CT (ncCT) and contrast-enhanced CT (ceCT) scans, the SegRap2025 challenge aims to enhance the generalizability and robustness of segmentation models across imaging centers and modalities. SegRap2025 comprises two tasks: \textit{Task01} addresses GTV segmentation using paired CT from the SegRap2023 dataset, with an additional external testing set to evaluate cross-center generalization, and \textit{Task02} focuses on LN CTV segmentation using multi-center training data and an unseen external testing set, where each case contains paired CT scans or a single modality, emphasizing both cross-center and cross-modality robustness. This paper presents the challenge setup and provides a comprehensive analysis of the solutions submitted by ten participating teams. For GTV segmentation task, the top-performing models achieved average Dice Similarity Coefficient (DSC) of 74.61\% and 56.79\% on the internal and external testing cohorts, respectively. For LN CTV segmentation task, the highest average DSC values reached 60.24\%, 60.50\%, and 57.23\% on paired CT, ceCT-only, and ncCT-only subsets, respectively. SegRap2025 establishes a large-scale multi-center, multi-modality benchmark for evaluating the generalization and robustness in radiotherapy target segmentation, providing valuable insights toward clinically applicable automated radiotherapy planning systems. The benchmark is available at: https://hilab-git.github.io/SegRap2025\_Challenge.

\end{abstract}


\end{frontmatter}


\section{Introduction}
Nasopharyngeal Carcinoma (NPC) is a malignant tumor with high prevalence in Southeast Asia and North Africa~\citep{lee2015management,chua2016nasopharyngeal,sun2019association}. Radiotherapy is the primary treatment for NPC. In the era of Intensity-Modulated Radiation Therapy (IMRT), accurate delineation of radiotherapy target volumes, such as Gross Tumor Volumes (GTVs), Clinical Target Volumes (CTVs), and Organs-at-Risk (OARs), is essential to ensure that high-dose radiation precisely covers tumor regions while sparing adjacent normal tissues~\citep{tang2019clinically}. Inaccurate delineation may result in underdosage of tumor areas or overexposure of critical organs, thereby compromising tumor control and increasing radiation-induced toxicity~\citep{wang2021guidelines,lin2019deep}. Manual delineation of these radiotherapy target volumes is highly labor-intensive and prone to inter-observer variability, motivating the development of automatic segmentation algorithms that can support clinical workflows~\citep{luo2025generalizable}.

Though deep learning has been widely used for medical image segmentation, the segmentation for radiotherapy from CT scans remains challenging due to low contrast, ambiguous boundaries, and large shape variation of OARs, GTVs and CTVs~\citep{ye2022comprehensive,liao2022automatic,luo2024multicenter}. Existing benchmarks~\citep{ang2014randomized,vallieres2017radiomics,raudaschl2017evaluation,oreiller2022head} are often constrained by limited sample sizes and data diversity, hindering the development of clinical applicable models and comprehensive evaluation. To address these, the SegRap2023 Challenge~\citep{luo2025segrap2023} established the first large-scale benchmark for NPC radiotherapy segmentation. The challenge provided paired non-contrast Computed Tomography (ncCT) and contrast-enhanced Computed Tomography (ceCT) scans from 200 patients collected at a single center, each annotated by expert radiation oncologists with 45 OARs and 2 GTVs. The challenge results demonstrated that automatic methods achieved high accuracy for large OARs, with the best-performing solutions reaching an average Dice Similarity Coefficient (DSC) of 86.70\%. In contrast, GTV segmentation remained suboptimal due to tumor heterogeneity and ambiguous boundaries.

Despite the progress achieved in SegRap2023, several key challenges limit the clinical deployment of automatic segmentation algorithms. First, robust cross-domain generalization is essential for real-world model deployment~\citep{wang2024dual}. Although many methods exhibit strong in-domain performance, their generalization ability to out-of-domain data, such as data from different imaging centers, scanners, or acquisition protocols, remains insufficient unexplored. Second, most existing approaches~\citep{lin2019deep,liao2022automatic,wang2024head,luo2025segrap2023} rely on Magnetic Resonance Imaging (MRI) or paired ncCT and ceCT scans, which are not consistently available in clinical practice due to equipment limitations, patient tolerance, or time constraints. Thus, the missing-modality problem requires further investigation. Third, existing benchmarks primarily adopt fully supervised learning paradigms and paid limited attention to data-efficient strategies for leveraging large-scale unlabeled clinical data. Moreover, while substantial efforts have focused on OAR and GTV segmentation, a comprehensive benchmark for Lymph Node CTV (LN CTV) segmentation is still lacking. The LN CTV plays an important role in prophylactic irradiation of NPC, as the region main contain cancer cells that are not visible in the image. Inaccurate segmentation of the LN CTV may lead to recurrence of the cancer. Consequently, developing segmentation algorithms for radiotherapy that generalizes robustly across imaging centers and modalities while effectively utilizing both labeled and unlabeled data remains an open and clinically significant challenge, and a benchmark for LN CTV segmentation is desirable for more comprehensive development and validation of segmentation models for radiotherapy.

\begin{figure}[t]
    \centering    
    \includegraphics[width=\columnwidth]{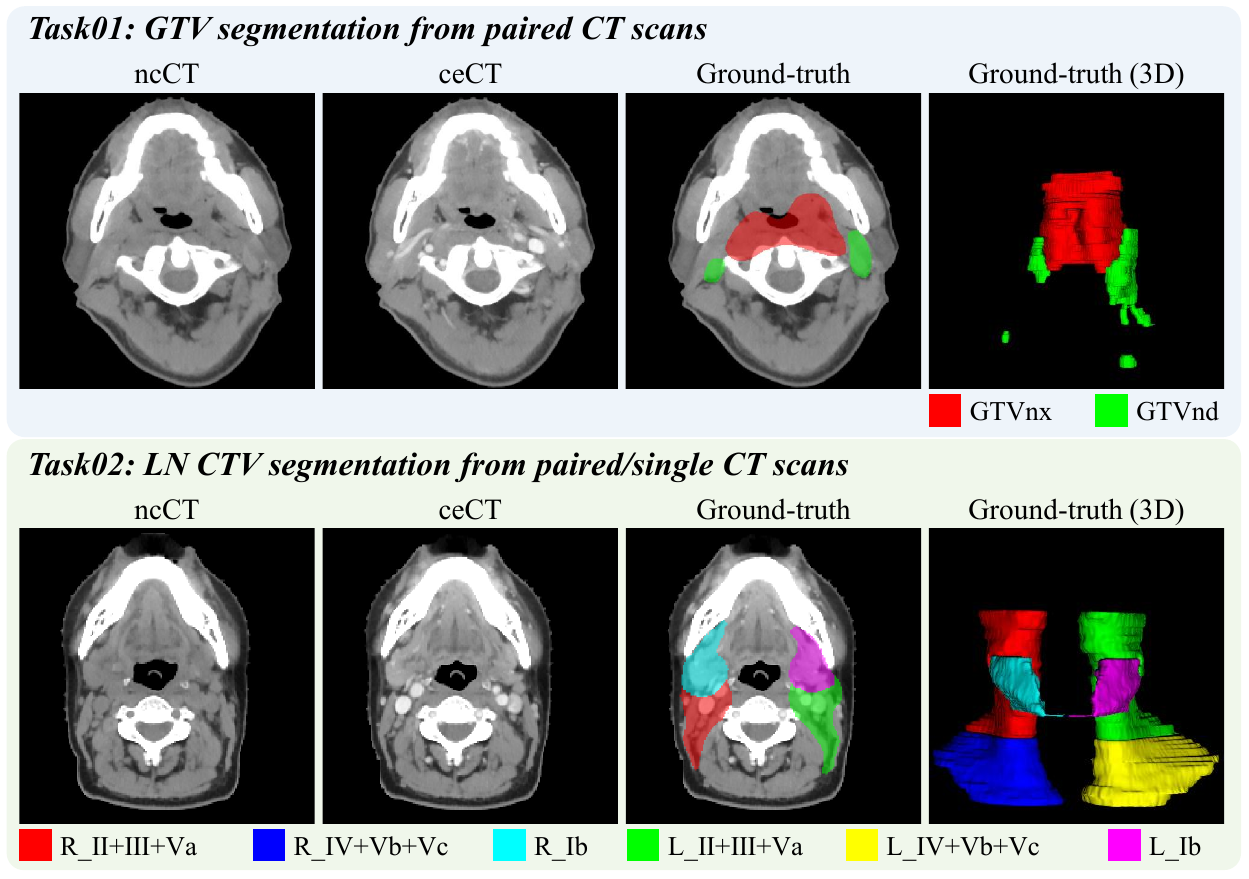}
    \caption{Overview of two sub-tasks in the SegRap2025 challenge.}
    \label{fig:overview}
\end{figure}

Building on the foundation of SegRap2023~\citep{luo2025segrap2023}, the SegRap2025 challenge was organized in conjunction with MICCAI 2025 to further advance model generalizability and modality robustness for NPC radiotherapy segmentation. The main contributions of SegRap2025 can be summarized as follows. \textit{First}, SegRap2025 evaluated the genralizability of algorithms for GTV segmentation (Task01), which reused the paired ncCT and ceCT training data from the SegRap2023 and extending the testing set with additional cases from an unseen external imaging center. \textit{Second}, SegRap2025 introduced a new task focusing on the segmentation of six LN CTVs (Task02). This task utilized a multi-center dataset comprising a mixture of paired and single-modality CT scans from four training centers and one unseen testing center, thereby filling the gap in current radiotherapy benchmarks. \textit{Third}, the challenge provided a large-scale unlabeled dataset, motivating the development of data-efficient segmentation models to improve model performance. \textit{Finally}, through its multi-center and multi-modality design, SegRap2025 enabled a comprehensive evaluation of model robustness under clinically relevant distribution shifts and missing-modality scenarios, facilitating their deployment in real-world radiotherapy environments.

\begin{table*}[t]
  \centering
  \caption{Summary of publicly available Gross Tumor Volume (GTV) segmentation datasets. ncCT and ceCT denote non-contrast and contrast-enhanced Computed Tomography (CT), respectively. GTVp and GTVnd represent primary and lymph node GTVs, respectively. HNC denotes head-and-neck cancer, and NPC denotes nasopharyngeal carcinoma.}
  \resizebox{\textwidth}{!}{
    \begin{tabular}{cccccccc}
    \hline
    Dataset & Site  & \makecell[c]{No. of\\Institutions} & Modality & \makecell[c]{No. of Cases\\(Training/Testing)} & Annotation & GTV types & Link \bigstrut\\
    \hline
    HECTOR2022 & HNC & 7     & FDG-PET, CT & 883 (524/359) & Full  & GTVp, GTVnd & \url{https://hecktor25.grand-challenge.org/} \bigstrut[t]\\
    HNTS-MRG2024 & HNC & 1     & T2w & 200 (150/50) & Full  & GTVp, GTVnd & \url{https://hntsmrg24.grand-challenge.org/} \\
    StructSeg2019 & NPC & 1     & ncCT  & 60 (50/10) & Full  & GTVp  & \url{https://structseg2019.grand-challenge.org} \\
    LNQ2023 & Chest & 1     & ceCT  & 513 (413/100) & Partial & GTVnd & \url{https://lnq2023.grand-challenge.org/} \\
    SegRap2023 & NPC & 1     & ncCT, ceCT & 200 (140/60) & Full  & GTVp, GTVnd & \url{https://segrap2023.grand-challenge.org} \\
    \hline
    \rowcolor{gray!10}
    SegRap2025 & NPC & 2 & \makecell[c]{ncCT, ceCT\\ncCT / ceCT} & \makecell[c]{260 (140/120)\\500 (500/0)} & \makecell[c]{Full\\Unlabeled} & \makecell[c]{GTVp, GTVnd\\-} & \url{https://hilab-git.github.io/SegRap2025\_Challenge} \bigstrut \\
    \hline
    \end{tabular}
    }
  \label{tab:datasets_summary}
\end{table*}

In this paper, we present the organization and results of the SegRap2025 Challenge. We first describe the dataset construction, annotation protocol, and challenge design. Then, we summarize the participating methods and evaluation framework, followed by an in-depth analysis of model performance across centers and modalities. Finally, we discuss key findings, remaining challenges, and insights revealed by the competition.

\section{{Related} works}
\subsection{GTV segmentation datasets and benchmarks}
As summarized in Table~\ref{tab:datasets_summary}, several public datasets have been released to support research on GTV segmentation for Head and Neck Cancers (HNC). The HECTOR challenge~\citep{oreiller2022head}, organized in conjunction with MICCAI for four years, provides FDG-PET/CT scans with manual annotations of both the primary GTV (GTVp, also named GTVnx) and lymph node GTV (GTVnd). The dataset has grown from 254 patients from four centers in HECKTOR2020 to over 1200 patients from 10 centers in HECKTOR2025, making it the largest benchmark for PET/CT-based HNC tumor segmentation. Similarly, the HNTS-MRG2024 Challenge~\citep{wahid2024overview} offers pre- and mid-radiotherapy T2-weighted MRI scans of 200 HNC patients with GTVp and GTVnd annotations, focusing on adaptive radiotherapy and longitudinal tumor response analysis. However, both HECKTOR and HNTS-MRG2024 primarily target general HNC rather than NPC, and their reliance on PET/CT or MRI limits their direct applicability to CT-based NPC radiotherapy planning.

For CT-based GTV segmentation, the StructSeg2019 Challenge provides ncCT scans of 60 patients with GTVp annotations, and the LNQ2023 Challenge~\citep{dorent2024lnq} offers 513 ceCT scans with partially annotated GTVnd masks for weakly supervised lymph node segmentation. Nevertheless, the small dataset size and incomplete annotations restrict their utility for training and evaluating clinically deployable models. To address this gap, our previous SegRap2023 Challenge~\citep{luo2025segrap2023} introduced the first large-scale benchmark for NPC radiotherapy segmentation based on paired ncCT and ceCT scans. It included 200 patients with expert manual annotations of GTVp and GTVnd, providing a valuable dataset for developing and validating deep learning-based models. While SegRap2023 significantly advanced CT-based NPC segmentation research, both the training and testing data originated from a single imaging center, limiting the evaluation of model generalization across institutions. Furthermore, existing benchmarks~\citep{dorent2024lnq,luo2025segrap2023} have been primarily restricted to fully supervised learning paradigms.

\subsection{LN CTV segmentation datasets and benchmarks}
Unlike visible tumors, LN CTV represents anatomical regions at risk of microscopic disease spread, which cannot be directly observed on imaging. Therefore, the delineation of LN CTVs relies heavily on anatomical knowledge and expert consensus. To reduce manual workload and mitigate inter- and intra-observer variability, several studies~\citep{van2020deep,strijbis2022deep,cardenas2021generating} have investigated deep learning methods for automatic segmentation of LN levels using CT scans from 70–85 HNC patients. However, these studies followed the 2013 updated consensus guideline~\citep{gregoire2014delineation}, which was primarily designed for general HNC cases and is not directly applicable to NPC, whose nodal metastatic patterns differ significantly from those in other HNC subtypes.

For NPC-specific LN CTV delineation,~\citet{lin2018delineation} proposed a dedicated LN CTV delineation map based on the spatial distribution of 10,651 metastatic lymph nodes and observed that several boundaries, particularly for levels Ib, II, IVa, and V, required adjustment relative to the 2013 guideline. Similarly,~\cite{zhao2022level} suggested additional modifications for levels IIb and V. These studies highlight the necessity for automatic segmentation models specifically tailored for NPC following the updated delineation protocols. However, these datasets have remained private, preventing systematic evaluation of how these models generalize across different imaging centers, scanners, and protocols. To address these gaps, we introduce the first comprehensive benchmark for NPC LN CTV segmentation. This benchmark is designed to evaluate model performance under multi-center and multi-modality distribution shifts, and encourages the development of automatic segmentation algorithms to bridge the gap between academic research and real-world clinical deployment.

\begin{table*}[t]
  \centering
  \caption{Training, validation, and testing dataset properties from all imaging centers for the SegRap2025 challenge. ceCT and ncCT represent contrast-enhanced computed tomography and non-contrast computed tomography, respectively. pts represents patients. $\checkmark$ and $\times$ denote labeled and unlabeled data, and $\sqrt{} \mkern-9mu {\smallsetminus}$ denotes labeled data but not available to participants.}
  \resizebox{\textwidth}{!}{
    \begin{tabular}{ccccccccccc}
    \hline
    \hline
    \multicolumn{2}{c}{Dataset} & Institution & Modalities & No. of pts & Labeled & Scanner & Thickness (mm) & In-plane resolution & Bulb voltage (kV) & Current (mA) \bigstrut\\
    \hline
    \hline
    \multirow{4}[2]{*}{Task01} & Training & SCH   & ncCT, ceCT & 120   & $\checkmark$  & Siemens & 3     & 512$\times$512 / 1024$\times$1024 & 120   & 300 \bigstrut[t]\\
          & Validation & SCH   & ncCT, ceCT & 20    & $\sqrt{} \mkern-9mu {\smallsetminus}$  & Siemens & 3     & 512$\times$512 / 1024$\times$1024 & 120   & 300 \\
          & \multirow{2}[1]{*}{Testing} & SCH   & ncCT, ceCT & 60    & $\sqrt{} \mkern-9mu {\smallsetminus}$     & Siemens & 3     & 512$\times$512 / 1024$\times$1024 & 120   & 300 \\
          &       & DHCJ & ncCT, ceCT & 60    & $\sqrt{} \mkern-9mu {\smallsetminus}$     & Siemens & 2.5   & 512$\times$512 & 120   & 200-250 \bigstrut[b]\\
    \hline
    \multirow{12}[6]{*}{Task02} & \multirow{6}[2]{*}{Training} & SPH   & ceCT  & 60    & $\checkmark$     & Siemens & 3     & 512$\times$512 & 120-140 & 280-380 \bigstrut[t]\\
          &       & APH   & ncCT  & 32    & $\checkmark$     & Siemens & 3     & 512$\times$512 & 120-140 & 280-380 \\
          &       & SMU   & ceCT  & 12    & $\checkmark$     & Siemens & 3     & 512$\times$512 & 120-140 & 280-380 \\
          &       & SCH   & ncCT, ceCT & 150   & $\checkmark$     & Philips & 3     & 512$\times$512 & 120   & 275-375 \\
          &       & SCH   & ceCT  & 8     & $\checkmark$     & Philips & 3     & 512$\times$512 & 120   & 275-375 \bigstrut[b] \\
\cline{2-11} & \multirow{3}[2]{*}{Validation} & \multirow{3}[2]{*}{DHCJ} & ncCT, ceCT & 20    & $\sqrt{} \mkern-9mu {\smallsetminus}$     & Siemens & 2.5   & 512$\times$512 & 120   & 200-250 \bigstrut[t]\\
          &       &       & ncCT  & 10    & $\sqrt{} \mkern-9mu {\smallsetminus}$     & Siemens & 2.5   & 512$\times$512 & 120   & 200-250 \\
          &       &       & ceCT  & 10    & $\sqrt{} \mkern-9mu {\smallsetminus}$     & Siemens & 2.5   & 512$\times$512 & 120   & 200-250 \bigstrut[b]\\
\cline{2-11} & \multirow{3}[2]{*}{Testing} & \multirow{3}[2]{*}{DHCJ} & ncCT, ceCT & 20    & $\sqrt{} \mkern-9mu {\smallsetminus}$     & Siemens & 2.5   & 512$\times$512 & 120   & 200-250 \bigstrut[t]\\
          &       &       & ncCT  & 10    & $\sqrt{} \mkern-9mu {\smallsetminus}$     & Siemens & 2.5   & 512$\times$512 & 120   & 200-250 \\
          &       &       & ceCT  & 10    & $\sqrt{} \mkern-9mu {\smallsetminus}$     & Siemens & 2.5   & 512$\times$512 & 120   & 200-250 \bigstrut[b]\\
    \hline
    Auxiliary & Training & SCH   & ncCT/ceCT & 500   & $\times$     & Siemens & 3     & 512$\times$512 & 120   & 300 \bigstrut \\
    
    \hline
    \hline
    \end{tabular}
    }
  \label{tab:dataset_characteristics}
\end{table*}

\section{SegRap2025 challenge setup}
\subsection{Challenge organization}
To evaluate existing algorithms and promote the development of generalizable and modality-robust segmentation models, the SegRap2025 Challenge was organized in conjunction with MICCAI 2025. As shown in Fig.~\ref{fig:overview}, participants were required to submit fully automatic segmentation algorithms addressing two tasks: (1) \textbf{\textit{Task01}}: segmentation of two Gross Tumor Volumes (GTVs), GTVp and GTVnd, from paired ncCT and ceCT scans, and (2) \textbf{\textit{Task02}}: segmentation of six Lymph Node Clinical Target Volumes (LN CTVs) from paired CT or single-modality (ceCT-only or ncCT-only) scans, including left (L)\_Ib, L\_II+III+Va, L\_IV+Vb+Vc, right (R)\_Ib, R\_II+III+Va, and R\_IV+Vb+Vc. Participants were allowed to leverage publicly available foundation models (e.g., pre-trained backbones or encoders), and their usage was required to be explicitly documented in the algorithm description. However, the use of additional external training data was strictly prohibited to ensure a fair comparison.

The challenge was run through the Github platform and consisted of three stages: training, validation, and testing. During the \textit{training} phase, participants were granted access to the training dataset after signing an End-User License Agreement (EULA) to ensure ethical and compliant data usage. The dataset remains publicly available to the research community after the competition to support future research and reproducibility. The \textit{validation} phase was open from June 30th, 2025 to August 31st, 2025. Each team was allowed up to five submissions per task to tune and evaluate their models on the validation set. 
In the final \textit{testing} phase, the testing data remained private to guarantee a fair and unbiased comparison. Participants were required to submit both their algorithm description and Docker containers before August 31st, 2025, following the official tutorial\footnote{\url{https://github.com/HiLab-git/SegRap2025_Docker}}. This ensured reproducibility and standardized evaluation under identical computational environments. Each team could submit its Docker container once, and all submissions were executed locally by the organizers using a standardized hardware configuration (24 GB GPU memory and 64 GB CPU memory) and consistent software environment. 

Only algorithms whose inference time did not exceed three minutes per case were considered valid submissions and included in the final ranking. Segmentation performance was evaluated using two widely adopted open-source Python packages, Evalutils\footnote{\url{https://evalutils.readthedocs.io/en/latest}} and MedPy\footnote{\url{https://loli.github.io/medpy}}, ensuring transparent and reproducible metric computation. The final leaderboard and ranking results were announced during the MICCAI 2025 Challenge event, following a thorough review process. Teams that failed to provide technical reports or reproducible submissions were excluded from the final ranking. The top three teams in each task received certificates and prize awards. Organizers could participate but were not eligible for awards.

\subsection{Data description}
\subsubsection{Task01: GTV segmentation}
The SegRap205 GTV segmentation dataset comprised NPC CT scans collected from two imaging centers, where 200 cases derived from the SegRap2023 Challenge~\citep{luo2025segrap2023}. Specifically, the \textit{training} set was from Sichuan Cancer Hospital (SCH) and consisted of two parts: paired and pre-aligned ncCT and ceCT scans from 120 patients with pixel-level GTV annotations (identical to the training set of SegRap2023), and 500 additional unlabeled single-modality CT scans, provided to encourage the development of semi-supervised and self-supervised learning strategies. The \textit{validation} set consisted of paired ncCT and ceCT scans from 20 patients from SCH, also consistent with SegRap2023. In addition to the \textit{internal testing} set from SegRap2023 (60 patients from SCH), SegRap2025 introduced an \textit{external testing} cohort comprising 60 additional patients from the Daguan Hospital of Chengdu Jinjiang (DHCJ), enabling evaluation of model generalization across imaging centers. Detailed dataset characteristics, including scanner configurations and acquisition parameters, are summarized in Table~\ref{tab:dataset_characteristics}, and more clinical information can be found in the SegRap2023 challenge paper~\citep{luo2025segrap2023}.

The initial contours of GTVs were delineated using 
MIM software\footnote{\url{https://www.mimsoftware.com}} according to the latest Radiation Therapy Oncology Group (RTOG) delineation guidelines\footnote{\url{https://www.rtog.org}} by W. Liao (MD, with ten years of experience in oncology radiation
therapy) and S.C. Zhang (MD, with more than twenty years of experience in oncology radiation therapy), together with their expert clinical team. In routine clinical workflow, radiation oncologists manually delineate GTVs until the contours meet the standards required for radiotherapy planning. During the initial delineation process, additional imaging modalities, such as MRI and PET, were referenced to ensure anatomical accuracy. To ensure annotation consistency and quality, W. Liao and S.C. Zhang performed detailed inspection and refinement of all contours using ITK-SNAP~\citep{yushkevich2006user}.

\subsubsection{Task02: LN CTV segmentation}
The SegRap2025 LN CTV segmentation dataset comprised NPC CT scans collected from five imaging centers. As shown in Table~\ref{tab:dataset_characteristics}, the \textit{training} data were derived from our previously released multi-center LN CTV dataset~\citep{luo2024multicenter}, which contains 262 NPC patients from SCH, Sichuan Provincial People’s Hospital (SPH), Anhui Provincial Hospital (APH), and Southern Medical University (SMU). Among them, 150 patients have paired ncCT and ceCT scans, while the remaining cases include only single-modality CT scans (ncCT-only or ceCT-only). This training set covers diverse imaging protocols, disease stages, and treatment strategies, as detailed in~\cite{luo2024multicenter}. Similar to \textit{Task01}, an additional 500 unlabeled CT scans were provided to facilitate the development of semi-supervised and self-supervised learning strategies. 

The \textit{validation} and \textit{testing} data were collected from an external cohort, DHCJ, to evaluate cross-center and cross-modality generalization. The validation set consists of 20 patients with paired ncCT and ceCT scans, 10 patients with ceCT-only scans, and 10 patients with ncCT-only scans. The testing set includes 40 patients with paired ncCT and ceCT scans, 30 patients with ceCT-only scans, and 30 patients with ncCT-only scans. 

The clinical annotation process was jointly conducted by a panel of three radiation oncologists, W. Liao, S.C. Zhang, and Y. Zhao, each with more than ten years of experience in head and neck radiotherapy. All cases were manually annotated with six basic LN CTVs based on clinical anatomical structures: right upper neck (R\_II+III+Va), right lower neck (R\_IV+Vb+Vc), right submandibular area (R\_Ib), left upper neck (L\_II+III+Va), left lower neck (L\_IV+Vb+Vc), and left submandibular area (L\_Ib). To ensure annotation consistency across centers, several representative cases were first delineated jointly as references before large-scale labeling. Pixel-wise LN CTV annotations for each center were delineated and verified by the expert panel using ITK-SNAP~\citep{yushkevich2006user} following the latest NPC-specific delineation guidelines to ensure high-quality ground truth.

\subsection{Evaluation}
Following the SegRap2023 challenge~\citep{luo2025segrap2023}, the SegRap2025 challenge used two widely-used metrics to comprehensively assess segmentation performance in terms of region overlap and boundary distance: Dice Similarity Coefficient (DSC) and Normalized Surface Dice (NSD). Both metrics range from 0 to 1, where higher values indicate better agreement between predictions and ground truth.
\begin{equation}
DSC(P, Y) = \frac{{2} \left | V_P \cap V_Y  \right | }{\left | V_P \right | + \left | V_Y \right |  },
\label{eq:DSC}
\end{equation}

\begin{equation}
NSD(P, Y) = \frac{\left | S_P \cap S^{(\tau)}_Y \right |  + \left | S_Y \cap S^{(\tau)}_P \right |}{\left | S_P \right | + \left | S_Y \right | }.
\label{eq:NSD}
\end{equation}
Where $V_P$ and $V_Y$ denote the predicted segmentation results and the ground truth, respectively. $S_P$ and $S_Y$ represent the surface voxels of the prediction and ground truth, and $S^{(\tau)}_P$ and $S^{(\tau)}_Y$ correspond to subsets of those surface voxels whose nearest-neighbour distance are within a tolerance threshold $\tau$. In the SegRap2025 challenge, the tolerance $\tau$ was set to 1 $mm$ for both GTVp and GTVnd, and 2 $mm$ for all six LN CTVs.

\subsection{Ranking}
The ranking rule was the same as in the SegRap2023 challenge~\citep{luo2025segrap2023}. For each test case, both DSC and NSD metrics were computed for every target class. In cases where a target structure was missing from the predicted segmentation, the corresponding metric values were assigned as 0. The mean DSC and NSD of each class were then averaged across all testing cases. Following the ranking rule in~\cite{bakas2018identifying}, teams were ranked separately for each class and each metric. The final ranking for each team was obtained by averaging its ranks across all classes and metrics within a task. This rule ensured a balanced and comprehensive evaluation, reflecting both volumetric accuracy and boundary precision across all anatomical targets.

\section{Overview of participating methods}\label{sec:set4}
A total of 48 teams registered for Task01, and 47 teams for Task02. During the final testing phase, both tasks had 10 teams that successfully submitted the containerized algorithms and detailed description of their methods. In this section, we summarize the methods employed by the participating teams for the two tasks. Table~\ref{tab:benchmark_task01} and Table~\ref{tab:benchmark_task02} summarize the key techniques of benchmarked algorithms, while Table~\ref{tab:Details_task01} and Table~\ref{tab:Details_task02} summarize the training details of benchmarked algorithms for Task01 and Task02, respectively.

\begin{table*}[t]
\setlength\tabcolsep{4pt}
  \centering
  \caption{Summary of the benchmarked algorithms for Task01. IN means intensity normalization. CC means Connected component-based post-processing.}
  \resizebox{\textwidth}{!}{
    \begin{tabular}{llcclc}
    \hline
    Team  & Pre-processing & Unlabeled Data & Public Weight & Data Augmentation (Training Time) & Post-processing \bigstrut\\
    \hline
    Space & Crop, IN, Resample & $\checkmark$ & $\times$ & Rotation, Scaling, Mirroring, Elastic, Gaussian noise \& blur & None \bigstrut[t]\\
    Mules and Horses & IN, Resample & $\times$ & $\checkmark$ & Rotation, Scaling, Mirroring, Gaussian noise \& blur, Contrast, Gamma & None \\
    FIT-Monash & Crop, IN, Resample & $\times$ & $\times$ & Rotation, Scaling, Mirroring, Elastic, Gaussian noise \& blur, Contrast, Gamma, Proposed LBGT & None \\
    17trophy & IN, Resample, Mixup & $\times$ & $\times$ & Rotation, Scaling, Mirroring, Gaussian noise \& blur, Contrast, Gamma & None \\
    alpinists & IN, Resample & $\times$ & $\times$ & Rotation, Scaling, Mirroring, Gaussian noise \& blur, Contrast, Gamma & None \\
    Winwin & Crop, IN, Resample & $\times$ & $\times$ & Rotation, Scaling, Mirroring, Gaussian noise \& blur, Contrast, Gamma, CutMix & None \\
    cemrg & IN, Resample & $\checkmark$ & $\times$ & Rotation, Scaling, Mirroring, Gaussian noise \& blur & None \\
    NPU\_SAIIP & Crop, IN, Resample & $\times$ & $\times$ & Rotation, Scaling, Gaussian noise \& blur, Contrast, Gamma & CC \\
    VIP-LAB & Crop, IN, Resample & $\times$ & $\checkmark$ & Rotation, Scaling, Mirroring, Gaussian noise \& blur, Contrast, Gamma & None \\
    junqiangmler & IN, Resample & $\times$ & $\times$ & Rotation, Scaling, Mirroring, Gaussian noise & CC \bigstrut[b]\\
    \hline
    Baseline & IN, Resample & $\times$ & $\times$ & Rotation, Scaling, Mirroring, Gaussian noise \& blur, Contrast, Gamma & None \\
    \hline
    \end{tabular}
    }
  \label{tab:benchmark_task01}
\end{table*}

\begin{table*}[t]
\setlength\tabcolsep{4pt}
  \centering
  \caption{Summary of the benchmarked algorithms for Task02. IN means intensity normalization. CC means Connected component-based post-processing.}
    \resizebox{\textwidth}{!}{
    \begin{tabular}{llcclc}
    \hline
    Team  & Pre-processing & Unlabeled Data & Public Weight & Data augmentation & Post-processing \bigstrut\\
    \hline
    Space & IN, Resample & $\times$ & $\times$ & Rotation, Scaling, Mirroring, Elastic, Gaussian noise \& blur & None \bigstrut[t]\\
    17trophy & IN, Resample, Mixup & $\times$ & $\times$ & Rotation, Scaling, Mirroring, Gaussian noise \& blur, Contrast, Gamma & None \\
    Mules and Horses & IN, Resample & $\times$ & $\times$ & Rotation, Scaling, Gaussian noise \& blur, Contrast, Gamma & None \\
    FIT-Monash & Crop, IN, Resample & $\times$ & $\times$ & Rotation, Scaling, Mirroring, Elastic, Gaussian noise \& blur, Contrast, Gamma, Proposed LBGT & None \\
    NPU\_SAIIP & Crop, IN, Resample & $\times$ & $\times$ & Rotation, Scaling, Gaussian noise \& blur, Contrast, Gamma & CC \\
    Winwin & IN, Resample & $\times$ & $\times$ & Rotation, Scaling, Gaussian noise \& blur, Contrast, Gamma & None \\
    junqiangmler & IN, Resample & $\times$ & $\times$ & Rotation, Scaling, Mirroring, Gaussian noise & CC \\
    SJTU\_LAB426 & Crop, IN, Resample & $\times$ & $\checkmark$ & Rotation, Scaling, Mirroring, Gaussian noise \& blur, Contrast, Gamma & None \\
    VIP-LAB & Crop, IN, Resample & $\times$ & $\checkmark$ & Rotation, Scaling, Mirroring, Gaussian noise \& blur, Contrast, Gamma & None \\
    cemrg & Crop, IN, Resample & $\times$ & $\times$ & Rotation, Scaling, Mirroring, Gaussian noise \& blur & None \bigstrut[b]\\
    \hline
    Baseline & IN, Resample & $\times$ & $\times$ & Rotation, Scaling, Gaussian noise \& blur, Contrast, Gamma & None \\
    \hline
    \end{tabular}
    }
  \label{tab:benchmark_task02}
\end{table*}

\begin{table*}[t]
\setlength\tabcolsep{4pt}
\rowcolors{2}{white}{gray!10}
  \centering
  \caption{Network architectures and training details of the benchmarked algorithms for Task01. CE and BCE mean cross-entropy and binary cross-entropy, respectively. × (*) refers to the number of ensemble models. ARSL represents adaptive region-specific loss.}
  \resizebox{\textwidth}{!}{
    \begin{tabular}{lcccccccl}
    \hline
    Team  & Backbone & Ensemble (Size) & Batch Size & Patch Size & Loss Function & Optimizer & Learning Rate & Device \bigstrut\\
    \hline
    Space & BLU-Net & × (5) & 4 & 56×192×160 & Dice and CE & AdamW & 0.0001 & NVIDIA GeForce RTX 5090 \bigstrut[t]\\
    Mules and Horses & UNet & × (5) & 2 & 64×192×192 & Dice and CE & Adam & 0.001 & NVIDIA GeForce RTX 3090 \\
    FIT-Monash & UNet & × (5) & 2 & \makecell[c]{64×192×192\\28×256×256} & Dice and CE & SGD & 0.01 & NVIDIA L40S \\
    17trophy & UNet & × (5) & 2 & 64×192×192 & Dice and CE & SGD & 0.01 & NVIDIA GeForce RTX 3090 \\
    alpinists & UNet & × (3) & 2 & 48×448×384 & Dice and CE & SGD & 0.01 & NVIDIA L40S \\
    Winwin & UNet & × (1) & 2 & 40×160×288  & ARSL, Dice and CE & SGD & 0.01  & NVIDIA GeForce RTX 2080 Ti \\
    cemrg & xLSTM-UNet & × (5) & 2 & 128×128×128 & Dice and CE & AdamW & 0.0002 & NVIDIA RTX A6000 \\
    NPU\_SAIIP & MedNeXt & × (3) & 2 & \makecell[c]{128×128×128\\64×192×192\\32×256×256} & Dice and CE  & AdamW   & 0.001  & NVIDIA GeForce RTX 3090 \\
    VIP-LAB & STUNet & × (5) & 2 & \makecell[c]{64×256×256\\128×128×128} & Dice and CE & SGD & 0.01 & NVIDIA GeForce RTX 4090 \\
    junqiangmler & VNet & × (1) & 4 & 256×256×256 & Dice and CE & AdamW & 0.0001 & NVIDIA A800 \bigstrut[b]\\
    \hline
    Baseline & nnUNet & × (1) & 2 & 28×256×256 & Dice and CE & SGD & 0.01 & NVIDIA GeForce RTX 2080 Ti \\
    \hline
    \end{tabular}
    }
  \label{tab:Details_task01}
\end{table*}

\begin{table*}[t]
\setlength\tabcolsep{4pt}
\rowcolors{2}{white}{gray!10}
  \centering
  \caption{Network architectures and training details of the benchmarked algorithms for Task02. CE and BCE mean cross-entropy and binary cross-entropy, respectively. SE means Squeeze-and-Excitation. × (*) refers to the number of ensemble models.}
  \resizebox{\textwidth}{!}{
    \begin{tabular}{lcccccccl}
    \hline
    Team  & Backbone & Ensemble (size) & Batch size & Patch Size & Loss function & Optimizer & Learning rate & Strategy for missing modality  \bigstrut\\
    \hline    
    Space & BLU-Net & × (1) & 2, 4 & \makecell[c]{32×192×160\\56×192×160} & Dice and CE & AdamW & 0.0001 & Modality-specific experts \bigstrut[t]\\
    17trophy & UNet & × (5) & 2 & 64×192×192 & Dice and CE & SGD & 0.01 &  Mixup, ceCT priority \\ 
    Mules and Horses & UNet & × (1) & 2 & 48×192×192 & Dice and CE  & Adam & 0.01  & Modality-specific experts \\
    FIT-Monash & UNet & × (5) & 2 & \makecell[c]{64×192×192\\28×256×256} & Dice and CE  & SGD & 0.01 & \makecell[l]{Intensity augmentation,\\ceCT priority} \\
    NPU\_SAIIP & MedNeXt & × (1) & 2 & 128×128×128 & Dice and CE & AdamW & 0.001 & Modality-specific experts \\
    Winwin & UNet & × (1) & 2 & 48×192×192 & Dice and CE & SGD & 0.01  & Replication, intensity clip \\
    junqiangmler & VNet & × (1) & 4 & 256×256×256 & Dice and CE & AdamW & 0.0001 & None \\
    SJTU\_LAB426 & \makecell[c]{CircleGAN\\SAM-Med2D\\UNet\\ResUNet\\MedNeXt} & × (4) & \makecell[c]{1\\8\\2\\2\\2} & \makecell[c]{256×256\\512×512\\48×192×192\\96×160×352\\80×112×224} & \makecell[c]{CircleGAN loss\\Dice and CE\\Dice and CE\\Dice and CE\\Dice and CE} & Adam & \makecell[c]{0.0002\\0.001\\0.01\\0.01 \\0.001} & Generative model \\
    VIP-LAB & STUNet & × (5) & 2 & \makecell[c]{64×256×256\\128×128×128} & Dice and CE & SGD & 0.01 & None \\
    cemrg & xLSTM-UNet & × (1) & 2 & 128×128×128 & Dice and CE & Adam & 0.001 & None \bigstrut[b]\\
    \hline
    Baseline & nnUNet & × (1) & 2 & 48×192×192 & Dice and CE & SGD & 0.01 & Modality-specific experts \\
    \hline
    \end{tabular}
    }
  \label{tab:Details_task02}
\end{table*}

\subsection{Task01: GTVs segmentation}
As summarized in Table~\ref{tab:benchmark_task01}, most teams relied on the standard data augmentation strategies provided by the nnUNet framework~\citep{isensee2021nnu}. Only \textit{17trophy}, \textit{FIT-Monash} and \textit{Winwin} extended it with additional pre-processing/augmentation strategies, which employed mixup, intensity-based transformations and CutMix to enhance data diversity, respectively. Among all participating teams, only two teams exploited the unlabeled dataset through semi-supervised (\textit{Space}) or self-supervised (\textit{cemrg}) training. Additionally, only one team (\textit{Mules and Horses}) used a foundation model, while the other teams mainly employed extensions to the nnUNet architecture or integrated auxiliary modules to enhance segmentation performance. Nine out of ten participating teams integrated customized networks into the nnUNet framework~\citep{isensee2021nnu} due to its strong effectiveness for head-and-neck tumor segmentation. In terms of loss design, only one team (\textit{Winwin}) experimented with a loss function beyond the standard combination of Dice and CE loss. Furthermore, eight out of ten teams used model ensembling, typically across folds or input patch sizes, to improve model robustness.

($1^{st}$ place, \textit{Space})
Team \textit{Space} proposed the Bootstrapped Learning Unified Network (BLU-Net) within a semi-supervised learning framework. They first trained a vanilla nnUNet~\citep{isensee2021nnu} on the SegRap2023 OAR dataset~\citep{luo2025segrap2023} to generate cropping masks for both labeled and unlabeled data. Subsequently, BLU-Net, featuring a U-shaped encoder–decoder architecture equipped with hierarchical gated convolutions, was trained on the cropped labeled set to effectively capture high-order spatial dependencies. Then, BLU-Net was used to select the top 40\% most confident unlabeled samples, whose pseudo-labels were added to the training set to further refine performance.

($2^{nd}$ place, \textit{Mules and Horses})
Team \textit{Mules and Horses} fine-tuned the LN-Seg-FM foundation model~\citep{luo2025dynamic}, originally designed for head and neck lymph node segmentation. During fine-tuning, the paired ncCT and ceCT scans were treated as independent cases, enabling the model to leverage complementary multi-modal information for more effective feature representation. Since soft-tissue boundaries were consistently clearer in ceCT, only ceCT scans were employed during inference.

($3^{rd}$ place, \textit{FIT-Monash})
Team \textit{FIT-Monash} developed a two-stage segmentation framework enhanced by a Learnable Bezier Grayscale Transform (LBGT)~\citep{wu2025sam}. A low-resolution nnUNet was first trained to predict a coarse foreground mask, which was used to produce a tight bounding box for subsequent fine segmentation. Then, a full-resolution nnUNet with an LBGT module performed refined segmentation within the cropped region. LBGT generated three contrast-adapted channels from the original CT through a data-driven Bezier tone-mapping, producing a pseudo-multimodal input that enhanced the discrimination of low-contrast structures. Notably, only ceCT was used for training and testing, and final predictions were generated by majority voting across five-fold cross-validation models.

($4^{th}$ place, \textit{17trophy})
Team \textit{17trophy} employed a nnUNet framework~\citep{isensee2021nnu} and expanded the dataset by offline fusing the ncCT and ceCT into mixed CT scans via mixup strategy, enhancing data diversity. The ncCT, ceCT, and mixed CT scans were treated as separate ones during training. During inference, test-time augmentation was applied to enhance robustness, and only ceCT predictions from cross-validation models were ensembled to generate final segmentation results.

($5^{th}$ place, \textit{alpinists})
Team \textit{alpinists} utilized residual encoders within the nnUNet~\citep{isensee2024nnu} framework. They employed a 3D full-resolution configuration, using both ncCT and ceCT scans as multi-channel input. For inference, predictions from cross-validation folds were ensembled via softmax probability averaging.

($6^{th}$ place, \textit{Winwin}) Team \textit{Winwin} first cropped the image into the body region and then clipped the intensity values of ncCT and ceCT to [-600, 600] and [-1000, 1000] for pre-processing, respectively, following the top-performing solutions in the SegRap2023 challenge~\citep{luo2025segrap2023}. During training, random copy-paste augmentation was applied to enhance data diversity, and adaptive region-specific loss functions~\citep{chen2023adaptive} were combined with Dice and CE losses to emphasize key regions. During inference, the same pre-processing pipeline as in the training stage was adopted.

($7^{th}$ place, \textit{cemrg}) 
Team \textit{cemrg} proposed a bi-directional extended Long Short-Term Memory (xLSTM)-UNet, extending the UNet encoder and bottleneck with xLSTM modules to capture global contextual dependencies and long-range spatial relationships. The model was first pre-trained on the unlabeled set using Masked Autoencoders (MAE)~\citep{he2022masked}, followed by five-fold cross-validation training on the labeled set. Predictions from five-fold models were ensembled as final results.

($8^{th}$ place, \textit{NPU\_SAIIP}) 
Team \textit{NPU\_SAIIP} employed 3D MedNeXt-L~\citep{roy2023mednext} as backbone network. Models were trained using three distinct patch sizes: 128×128×128, 64×192×192, and 32×256×256, preserving complementary receptive-field information and improving robustness. During inference, the logits produced by the three patch-specific models were averaged to generate final predictions.

($9^{th}$ place, \textit{VIP-LAB})
Team \textit{VIP-LAB} employed a 3D STUNet~\citep{huang2023stu} for tumor and lymph node segmentation, leveraging its encoder-decoder structure with skip connections to capture long-range context and fine boundaries. A centered cropping strategy reduced input volumes to 128×512×512 to focus the network on the regions of interest. During inference, model ensembling was across folds and test-time augmentation were applied to enhance generalization and boundary consistency.

($10^{th}$ place, \textit{junqiangmler}) 
Team \textit{junqiangmler} utilized the 3D VNet architecture~\citep{milletari2016v} with Z-score normalization and resizing into fixed size of 256×256×256. During inference, small connected component were removed as post-processing.

(\textit{Baseline}) To verify if the participants' methods bring progress from the standard training method, we implemented a baseline method that utilizes the nnUNet~\citep{isensee2021nnu} backbone with the default 3D full-resolution configuration.

\begin{table*}[t]
  \centering
  \caption{Summary of the quantitative evaluation results of GTVp and GTVnd segmentation on the internal and external testing cohorts by the ten teams. * denotes p-value < 0.05 when comparing the best-performing team with the Baseline.}
  \resizebox{\textwidth}{!}{
    \begin{tabular}{cccc|ccc|ccc|ccc}
    \hline
    \multirow{3}[6]{*}{Method} & \multicolumn{6}{c|}{Internal testing cohort}  & \multicolumn{6}{c}{External testing cohort} \bigstrut\\
\cline{2-13}          & \multicolumn{3}{c|}{DSC (\%)} & \multicolumn{3}{c|}{NSD (\%)} & \multicolumn{3}{c|}{DSC (\%)} & \multicolumn{3}{c}{NSD (\%)} \bigstrut\\
\cline{2-13}          & GTVp  & GTVnd & Average & GTVp  & GTVnd & Average & GTVp  & GTVnd & Average & GTVp  & GTVnd & Average \bigstrut\\
    \hline
    Space & 77.96±7.37 & 69.57±12.64 & 73.76±8.41 & 35.82±12.17 & \textbf{64.90±15.23*} & \textbf{50.36±11.37*} & 49.43±14.29 & \textbf{63.57±13.46*} & 56.50±11.34 & 12.11±8.85 & 56.81±16.61 & 34.46±10.28 \bigstrut[t]\\
    Mules and Horses & \textbf{79.47±6.62*} & \textbf{69.74±11.83*} & \textbf{74.60±7.64*} & \textbf{36.56±10.51*} & 64.06±14.16 & 50.31±10.58 & 47.22±13.76 & 62.86±14.04 & 55.04±11.77 & 9.33±7.56 & 55.62±18.41 & 32.48±10.65 \\
    FIT-Monash & 77.38±7.75 & 67.96±12.88 & 72.67±8.62 & 34.37±11.53 & 63.51±15.59 & 48.94±11.55 & 50.34±14.33 & 63.23±13.85 & \textbf{56.78±11.26*} & 12.71±9.36 & \textbf{58.14±17.46*} & \textbf{35.43±10.85*} \\
    17trophy & 77.86±8.15 & 68.73±13.37 & 73.29±9.58 & 35.60±13.30 & 63.56±15.96 & 49.58±12.82 & \textbf{50.74±12.97*} & 60.08±14.48 & 55.41±10.74 & \textbf{14.54±6.76*} & 52.97±15.73 & 33.75±9.40 \\
    alpinists & 77.79±7.75 & 67.48±14.96 & 72.63±9.16 & 35.46±11.59 & 63.75±16.25 & 49.61±11.03 & 49.50±14.32 & 63.33±13.48 & 56.42±11.35 & 11.88±9.46 & 56.33±16.81 & 34.11±10.58 \\
    Winwin & 78.59±7.14 & 68.60±12.04 & 73.60±7.90 & 36.21±11.37 & 63.59±14.58 & 49.90±10.65 & 48.75±14.27 & 61.79±13.80 & 55.27±11.53 & 10.84±8.39 & 53.90±17.47 & 32.37±10.30 \\
    cemrg & 78.38±7.40 & 67.72±12.60 & 73.05±8.23 & 36.54±12.26 & 62.20±14.56 & 49.37±11.12 & 47.91±13.32 & 62.06±13.74 & 54.99±11.41 & 10.42±7.76 & 54.85±17.55 & 32.64±10.55 \\
    NPU\_SAIIP & 77.75±7.70 & 66.30±15.64 & 72.02±9.31 & 35.04±11.37 & 61.38±16.45 & 48.21±11.05 & 49.65±13.76 & 62.04±13.06 & 55.84±10.80 & 11.72±8.63 & 54.62±16.72 & 33.17±10.14 \\
    VIP-LAB & 76.11±8.35 & 63.06±18.04 & 69.59±11.50 & 33.70±13.19 & 58.20±19.76 & 45.95±14.21 & 47.22±12.74 & 51.81±19.88 & 49.52±11.64 & 12.32±7.66 & 42.39±15.69 & 27.36±8.47 \\
    junqiangmler & 67.84±9.69 & 40.11±14.37 & 53.98±10.09 & 22.82±8.88 & 32.30±11.20 & 27.56±7.84 & 42.67±14.77 & 45.59±13.69 & 44.13±10.12 & 10.36±8.20 & 38.07±9.63 & 24.21±6.45 \bigstrut[b]\\
    \hline
    Baseline & 77.53±7.16 & 66.38±14.38 & 71.95±8.55 & 33.04±10.04 & 61.96±15.47 & 47.50±10.22 & 48.19±13.85 & 61.19±14.10 & 54.69±11.25 & 10.91±8.32 & 53.88±17.98 & 32.39±10.66 \bigstrut\\
    \hline
    \end{tabular}
    }
  \label{tab:GTV_DSC_NSD}
\end{table*}

\begin{table}[t]
  \centering
  \caption{Rankings of methods in terms of DSC and NSD scores for GTV segmentation.}
  \resizebox{\columnwidth}{!}{
    \begin{tabular}{ccc|cc|cc|ccc}
    \hline
    \multirow{3}[6]{*}{Method} & \multicolumn{4}{c|}{Internal testing cohort} & \multicolumn{4}{c}{External testing cohort} & \multirow{3}[6]{*}{\makecell[c]{Overall\\Rank}} \bigstrut\\
\cline{2-9}          & \multicolumn{2}{c|}{DSC Rank} & \multicolumn{2}{c|}{NSD Rank} & \multicolumn{2}{c|}{DSC Rank} & \multicolumn{2}{c}{NSD Rank} &  \bigstrut\\
\cline{2-9}          & GTVp  & GTVnd & GTVp & GTVnd & GTVp  & GTVnd & GTVp  & GTVnd &  \bigstrut\\
    \hline
    Space & 4     & 2     & 4     & 1     & 5     & 1     & 4     & 2     & 1 \bigstrut[t]\\
    Mules and Horses & 1     & 1     & 1     & 2     & 8     & 4     & 10    & 4     & 2 \\
    FIT-Monash & 8     & 5     & 8     & 6     & 2     & 3     & 2     & 1     & 3 \\
    17trophy & 5     & 3     & 5     & 5     & 1     & 8     & 1     & 8     & 4 \\
    alpinists & 6     & 7     & 6     & 3     & 4     & 2     & 5     & 3     & 4 \\
    Winwin & 2     & 4     & 3     & 4     & 6     & 7     & 7     & 7     & 6 \\
    cemrg & 3     & 6     & 2     & 7     & 7     & 5     & 8     & 5     & 7 \\
    NPU\_SAIIP & 7     & 8     & 7     & 8     & 3     & 6     & 6     & 6     & 8 \\
    VIP-LAB & 9     & 9     & 9     & 9     & 8     & 9     & 3     & 9     & 9 \\
    junqiangmler & 10    & 10    & 10    & 10    & 10    & 10    & 9     & 10    & 10 \bigstrut[b]\\
    \hline
    \end{tabular}
    }
  \label{tab:GTV_Rank}
\end{table}

\subsection{Task02: LN CTVs segmentation}
Most teams also built their solutions on the nnUNet~\citep{isensee2021nnu} framework with some customized network designs. Missing-modality handling was a central challenge in Task02, and the participants explored diverse strategies, including: cross-modality knowledge transfer from the paired CT to ncCT-only or ceCT-only (\textit{Space}), expert models for each modality (\textit{NPU\_SAIIP}), modality synthesis using a generative model (\textit{SJTU\_LAB426}), and using ceCT exclusively during inference (\textit{17trophy}, \textit{FIT-Monash}) due to its higher soft-tissue contrast. For teams whose architectures were largely identical to Task01, only Task02-specific modifications are described.

($1^{st}$ place, \textit{Space})
Team \textit{Space} used BLU-Net again for Task02, but without semi-supervised learning (i.e., no pseudo-labels were generated from unlabeled data). Instead, BLU-Net was pre-trained on the OAR dataset and then fine-tuned on paired CT scans for cross-modality knowledge transfer. The paired-CT model was subsequently fine-tuned to produce modality-specific experts for ncCT and ceCT, effectively addressing missing-modality issues during inference.

($2^{nd}$ place, \textit{17trophy})
Team \textit{17trophy} employed a nnUNet-based framework similar to their GTV segmentation pipeline and enhanced data diversity by generating mixed CT images through equal-weighted fusion on paired CT scans. The resulting ceCT, ncCT, and mixed CT scans were treated as separate samples. For inference, they combined predictions from cross-validation models with test-time augmentation to improve segmentation robustness. Notably, for cases with paired CT scans, only ceCT images were used during inference.

($3^{rd}$ place, \textit{Mules and Horses})
Team \textit{Mule and Horses} trained nnUNet~\citep{isensee2021nnu} from scratch on the heterogeneous multi-modal dataset, treating paired ncCT and ceCT scans as independent samples. During inference, single-modality cases relied on their respective predictions, where the predictions for paired cases were ensembled across ncCT and ceCT.

($4^{th}$ place, \textit{FIT-Monash})
Team \textit{FIT-Monash} reused their two-stage nnUNet combined with the LBGT framework~\citep{wu2025sam}. Coarse foreground masks were generated from a low-resolution nnUNet and then used to generate an expanded bounding box. The second stage performed refined segmentation using a full-resolution nnUNet on the cropped volume with the LBGT module. Similar to \textit{17trophy}, they prioritized ceCT scans during inference and used ncCT scans only when ceCT scans was unavailable.

($5^{th}$ place, \textit{NPU\_SAIIP})
Team \textit{NPU\_SAIIP} trained modality-specific expert models, ceCT-only and ncCT-only, using the 3D MedNeXt-L~\citep{roy2023mednext} backbone based on the nnUNet architecture~\citep{isensee2021nnu}. During inference, predictions from the corresponding expert model were aggregated through an ensembling strategy. For cases with paired CT scans, the logits from two expert models were averaged as the final results.

($6^{th}$ place, \textit{Winwin}) Team \textit{Winwin} trained a two-channel nnUNet~\citep{isensee2021nnu} to achieve the segmentation of LN CTVs, where the missing-modality CT scans were supplemented by copying data from the existing modality. Similar to the GTV segmentation, intensity values of ncCT and ceCT scans were clipped into [-600, 600] and [-1000, 1000], respectively. Given the symmetry of NPC LN CTVs, flip was disabled, and rotation angles restricted to [-30, 30]. The inference process followed identical preprocessing.

($7^{th}$ place, \textit{junqiangmler})
Team \textit{junqiangmler} reused the 3D VNet~\citep{milletari2016v} pipeline. The pre-processing and post-processing processes were the same as those used for GTV segmentation.

($8^{th}$ place, \textit{SJTU\_LAB426})
Team \textit{SJTU\_LAB426} addressed the missing-modality problem by integrating a generative model with a multi-architecture ensemble. The method first employed a 2D CircleGAN~\citep{shim2020circlegan} to synthesize the absent modality (either ncCT or ceCT) at the slice level, and then paired it with the available one, resulting in a dual-channel input that preserves complementary information. Subsequently, this complete dual-channel input is fed into a powerful multi-model ensemble consisting of four diverse segmentation backbones, including SAM-Med2D~\citep{cheng2023sammed2d} fine-tuned through LoRA~\citep{hu2022lora}, nnUNet, nnUNet with residual encoder, and 3D MedNeXt~\citep{roy2023mednext}. During inference, probability maps from the four models were averaged to generate the final segmentation results.

\begin{figure*}[t]
    \centering\includegraphics[width=\textwidth]{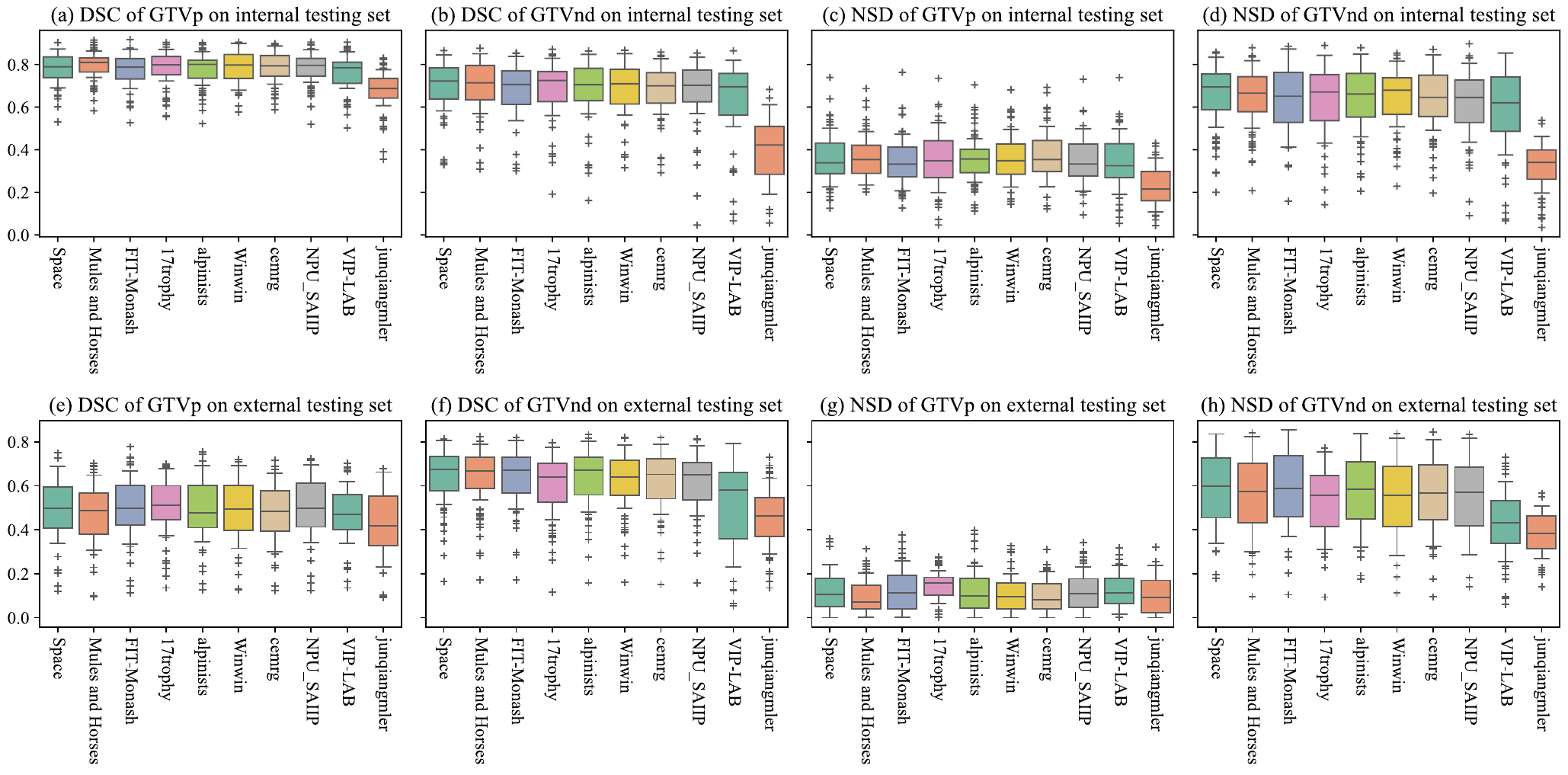}
    \caption{Boxplot of the patient-level average segmentation performance for GTVs in terms of DSC and NSD.}
    \label{fig:boxplot_task01}
\end{figure*}

\begin{figure}[t]
    \centering
    \includegraphics[width=\columnwidth]{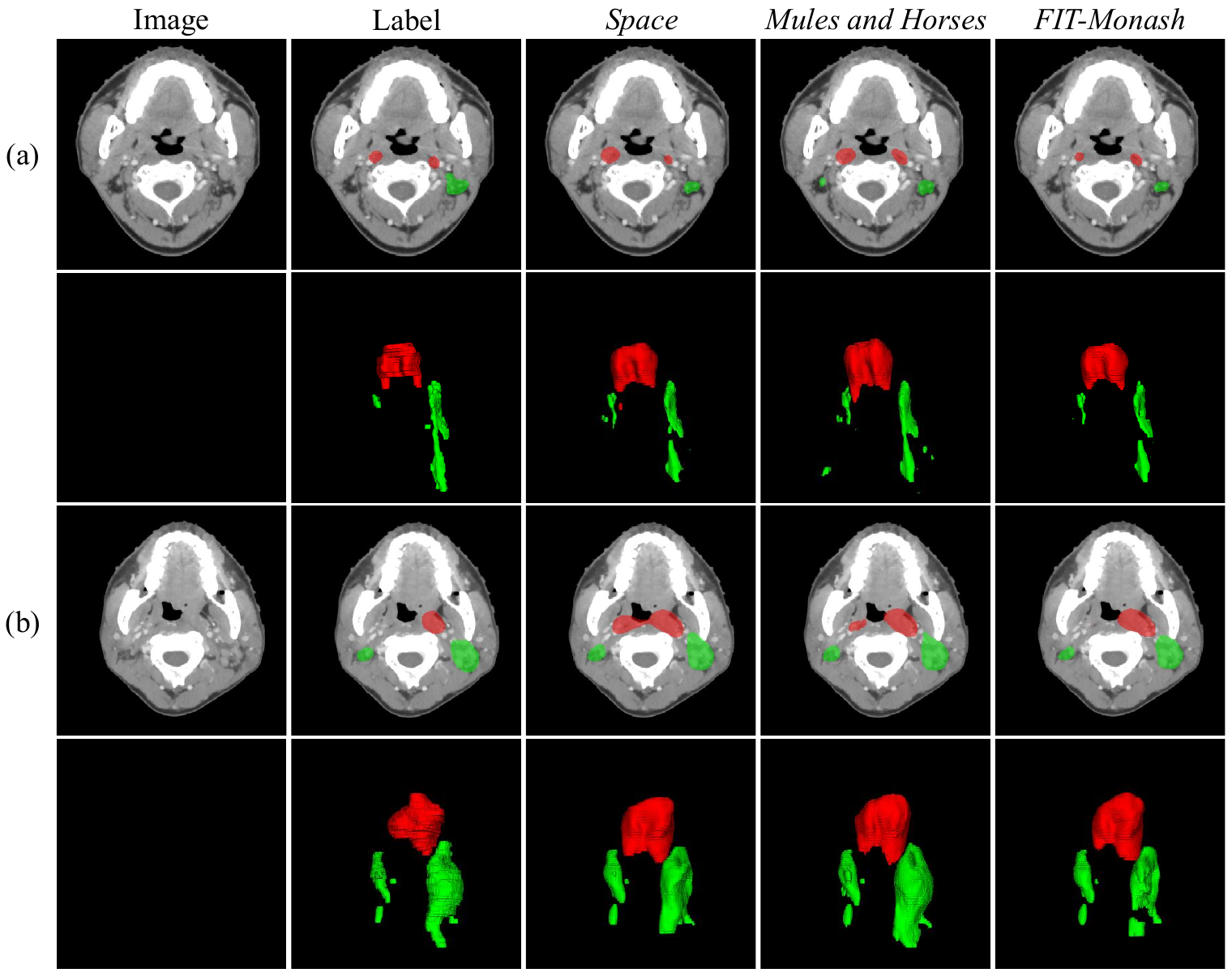}
    \caption{Qualitative GTV segmentation using the top three teams on the internal testing set. \textcolor{red}{Red} and \textcolor{green}{green} denote GTVp and GTVnd, respectively.}
    \label{fig:task01_vis_internal}
\end{figure}

\begin{figure}[t]
    \centering
    \includegraphics[width=\columnwidth]{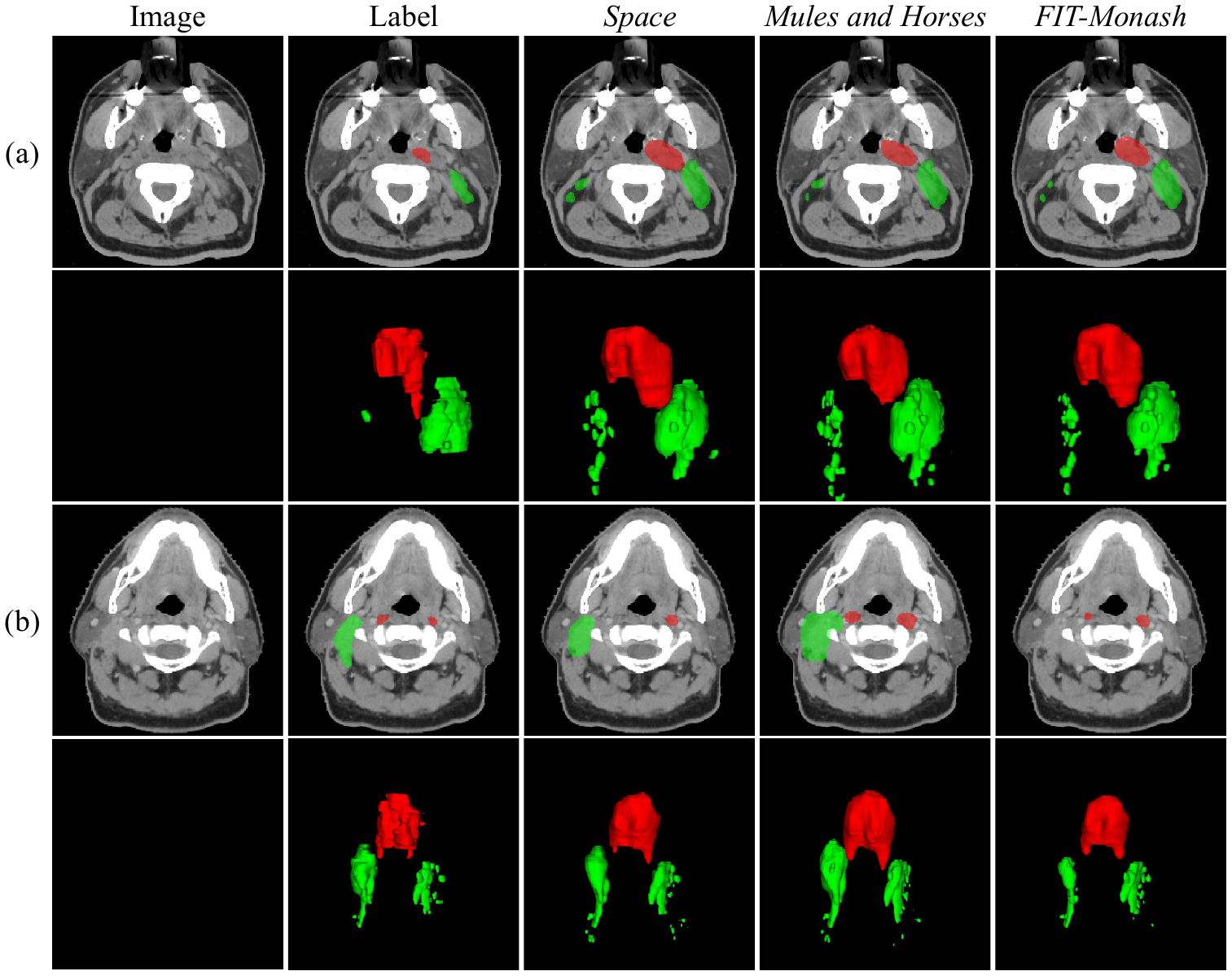}
    \caption{Qualitative GTV segmentation using the top three teams on the external testing set. \textcolor{red}{Red} and \textcolor{green}{green} denote GTVp and GTVnd, respectively.}
    \label{fig:task01_vis_external}
\end{figure}

($9^{th}$ place, \textit{VIP-LAB}) Team \textit{VIP-LAB} employed a 3D STUNet~\citep{huang2023stu} for the segmentation of LN CTVs, leveraging its encoder-decoder structure to capture long-range context and fine boundaries. Two random cropping scales of 128×128×128 and 64×256×256 were used to emphasize regions of interest. During inference, model ensembling was applied to enhance generalization and boundary consistency.

($10^{th}$ place, \textit{cemrg}) Team \textit{cemrg} adopted the bi-directional xLSTM-UNet again for LN CTV segmentation. Unlike GTV segmentation, the model was trained directly on the labeled training data, without pretraining on the unlabeled data.

(\textit{Baseline}) Similarly to the GTV segmentation task, to understand to what degree the participants' solutions improve the performance from a standard training method, we implemented a baseline method using the nnUNet~\citep{isensee2021nnu} backbone with the default 3D full-resolution configuration for each modality. During inference, probability maps from the two expert models were averaged as the final results for paired CT subset.

\begin{table*}[t]
\setlength\tabcolsep{3pt}
  \centering
  \caption{Rankings of methods in terms of DSC/NSD scores for LN CTV segmentation.}
  \resizebox{\textwidth}{!}{
    \begin{tabular}{ccccccc|cccccc|ccccccc}
    \hline
    \multirow{2}[2]{*}{Method} & \multicolumn{6}{c|}{Paired CT subset}         & \multicolumn{6}{c|}{ncCT-only subset}         & \multicolumn{6}{c}{ceCT-only subset}          & \multirow{2}[2]{*}{\makecell[c]{Overall\\Rank}} \bigstrut[t]\\
    \cline{2-19}          & R\_II+III+Va & L\_II+III+Va & R\_IV+Vb+Vc & L\_IV+Vb+Vc & R\_Ib & L\_Ib & R\_II+III+Va & L\_II+III+Va & R\_IV+Vb+Vc & L\_IV+Vb+Vc & R\_Ib & L\_Ib & R\_II+III+Va & L\_II+III+Va & R\_IV+Vb+Vc & L\_IV+Vb+Vc & R\_Ib & L\_Ib & \bigstrut \\
    \hline
    Space & 1/1   & 4/5   & 6/5   & 4/4   & 1/1   & 1/1   & 2/2   & 1/1   & 6/5   & 2/1   & 1/1   & 1/1   & 6/6   & 4/2   & 7/7   & 3/2   & 4/2   & 4/1   & 1 \bigstrut[t]\\
    17trophy & 2/2   & 1/1   & 2/2   & 3/5   & 7/6   & 6/6   & 1/1   & 2/2   & 1/1   & 5/6   & 6/5   & 6/5   & 2/2   & 1/2   & 1/1   & 1/5   & 3/3   & 3/3   & 2 \\
    Mules and Horses & 4/4   & 3/3   & 4/4   & 6/6   & 5/5   & 5/4   & 6/3   & 4/5   & 4/2   & 1/2   & 3/2   & 2/3   & 4/4   & 1/1   & 4/4   & 2/1   & 2/4   & 2/2   & 3 \\
    FIT-Monash & 3/3   & 2/2   & 1/2   & 1/3   & 6/7   & 4/5   & 4/5   & 5/6   & 3/4   & 3/3   & 5/3   & 4/4   & 3/3   & 3/4   & 3/3   & 4/3   & 5/5   & 5/4   & 4 \\
    NPU\_SAIIP & 5/5   & 5/4   & 5/6   & 5/7   & 8/8   & 8/8   & 3/4   & 6/4   & 5/3   & 6/5   & 4/4   & 7/7   & 1/1   & 5/5   & 2/2   & 5/4   & 6/8   & 6/7   & 5 \\
    Winwin & 6/8   & 6/6   & 3/1   & 2/1   & 2/2   & 2/2   & 7/8   & 3/3   & 2/6   & 4/4   & 8/8   & 5/6   & 7/7   & 6/7   & 8/8   & 8/7   & 7/7   & 8/8   & 6 \\
    junqiangmler & 7/6   & 7/7   & 8/8   & 7/2   & 3/4   & 7/7   & 8/7   & 8/8   & 8/7   & 8/7   & 2/6   & 8/8   & 5/5   & 7/6   & 5/5   & 7/6   & 1/1   & 1/5   & 7 \\
    SJTU\_LAB426 & 8/7   & 9/8   & 9/9   & 9/9   & 4/3   & 3/3   & 5/5   & 7/7   & 9/9   & 9/9   & 7/7   & 3/2   & 9/9   & 9/9   & 9/9   & 9/9   & 9/9   & 7/9   & 8 \\
    VIP-LAB & 9/9   & 8/9   & 7/7   & 8/8   & 9/9   & 9/9   & 9/9   & 9/9   & 7/8   & 7/8   & 9/9   & 9/9   & 8/8   & 8/8   & 6/6   & 6/8   & 8/6   & 9/6   & 9 \\
    cemrg & 10/10 & 10/10 & 10/10 & 10/10 & 10/10 & 10/10 & 10/10 & 10/10 & 10/10 & 10/10 & 10/10 & 10/10 & 10/10 & 10/10 & 10/10 & 10/10 & 10/10 & 10/10 & 10 \bigstrut[b]\\
    \hline
    \end{tabular}
    }
  \label{tab:LNCTV_Rank}
\end{table*}

\begin{table*}[t]
\setlength\tabcolsep{3pt}
  \centering
  \caption{Summary of the quantitative evaluation results of LN CTV segmentation by the ten teams. * denotes p-value < 0.05 when comparing the best-performing team with the Baseline.}
  \resizebox{\textwidth}{!}{
    \begin{tabular}{c|ccccccc|ccccccc}
    \hline
    \hline
    \multirow{3}[6]{*}{Method} & \multicolumn{14}{c}{Paired CT subset} \bigstrut\\
\cline{2-15}          & \multicolumn{7}{c|}{DSC (\%)}                 & \multicolumn{7}{c}{NSD (\%)} \bigstrut\\
\cline{2-15}          & R\_II+III+Va & L\_II+III+Va & R\_IV+Vb+Vc & L\_IV+Vb+Vc & R\_Ib & L\_Ib & Average & \multicolumn{1}{c}{R\_II+III+Va} & \multicolumn{1}{c}{L\_II+III+Va} & \multicolumn{1}{c}{R\_IV+Vb+Vc} & \multicolumn{1}{c}{L\_IV+Vb+Vc} & \multicolumn{1}{c}{R\_Ib} & \multicolumn{1}{c}{L\_Ib} & Average \bigstrut\\
    \hline
    Space & \textbf{78.36±5.62*} & 76.95±5.69 & 67.12±14.83 & 66.59±14.00 & \textbf{36.50±8.97*} & \textbf{35.89±7.75*} & \textbf{60.23±7.45*} & \textbf{58.20±12.51*} & 55.60±13.09 & 44.76±13.65 & 43.60±14.01 & \textbf{31.45±7.75*} & \textbf{31.54±5.66*} & \textbf{44.19±8.54*} \\
    17trophy & 77.97±5.80 & \textbf{77.45±5.51*} & 67.79±14.53 & 66.68±14.00 & 32.31±8.30 & 32.51±7.81 & 59.12±7.43 & 58.18±12.45 & \textbf{56.87±12.15*} & 45.36±14.67 & 43.11±15.06 & 27.84±6.59 & 28.59±6.76 & 43.32±8.81 \\
    Mules and Horses & 77.37±5.40 & 77.20±5.25 & 67.24±13.90 & 66.03±13.72 & 32.44±8.25 & 32.59±8.23 & 58.81±7.19 & 56.61±11.61 & 56.43±11.64 & 44.80±14.28 & 42.70±14.76 & 27.86±6.29 & 28.77±6.63 & 42.86±8.49 \\
    FIT-Monash & 77.95±5.60 & 77.38±5.65 & \textbf{67.99±13.09*} & \textbf{67.09±13.32*} & 32.43±7.93 & 32.60±8.01 & 59.24±7.09 & 57.71±12.02 & 56.79±12.81 & 45.36±14.06 & 43.64±14.50 & 27.61±5.89 & 28.65±5.83 & 43.29±8.40 \\
    NPU\_SAIIP & 77.36±5.85 & 76.66±5.63 & 67.14±12.99 & 66.09±13.25 & 31.90±8.84 & 31.19±8.37 & 58.39±6.99 & 56.48±12.37 & 55.76±12.59 & 43.84±13.77 & 42.44±13.92 & 26.51±6.27 & 26.95±6.27 & 42.00±8.15 \\
    Winwin & 76.62±5.51 & 76.14±5.64 & 67.49±14.65 & 66.86±13.39 & 35.08±8.56 & 33.64±7.96 & 59.30±7.31 & 55.36±11.07 & 54.59±12.40 & \textbf{45.92±14.12*} & \textbf{43.68±15.25*} & 29.99±6.84 & 29.80±6.05 & 43.22±8.54 \\
    junqiangmler & 76.25±6.07 & 74.85±6.75 & 64.19±16.54 & 64.87±17.38 & 34.84±7.87 & 32.44±7.43 & 57.91±8.04 & 56.16±11.27 & 53.92±12.97 & 42.12±12.83 & 43.66±14.10 & 27.87±6.08 & 27.95±5.76 & 41.95±7.69 \\
    SJTU\_LAB426 & 73.94±17.97 & 72.81±17.69 & 34.86±34.68 & 35.73±34.83 & 33.21±11.21 & 33.11±11.28 & 47.28±16.22 & 55.51±17.16 & 53.57±17.47 & 24.19±25.02 & 23.42±24.62 & 28.78±9.24 & 29.74±9.18 & 35.87±12.88 \\
    VIP-LAB & 73.60±8.97 & 72.96±7.49 & 66.16±13.86 & 64.57±15.38 & 28.19±8.47 & 28.43±8.21 & 55.65±7.59 & 53.29±13.61 & 52.00±13.98 & 42.88±13.53 & 41.27±15.71 & 25.58±6.40 & 25.78±7.42 & 40.13±8.96 \\
    cemrg & 45.73±9.32 & 39.22±13.03 & 21.96±15.81 & 11.24±11.57 & 9.83±6.06 & 10.36±6.89 & 23.06±6.97 & 22.88±6.04 & 20.29±7.27 & 12.28±8.79 & 5.52±5.94 & 12.03±4.09 & 14.35±6.16 & 14.56±4.19 \\
    \hline
    Baseline & 72.46±12.85 & 71.04±11.64 & 65.06±15.37 & 64.38±15.28 & 28.25±11.51 & 28.12±11.28 & 54.89±9.63 & 52.56±13.00 & 50.21±13.63 & 41.51±13.01 & 40.52±13.66 & 24.31±8.23 & 24.99±8.54 & 39.02±8.92 \\
    \hline
    \hline
    \multirow{3}[6]{*}{Method} & \multicolumn{14}{c}{ceCT-only subset} \bigstrut\\
\cline{2-15}          & \multicolumn{7}{c|}{DSC (\%)}                 & \multicolumn{7}{c}{NSD (\%)} \bigstrut\\
\cline{2-15}          & R\_II+III+Va & L\_II+III+Va & R\_IV+Vb+Vc & L\_IV+Vb+Vc & R\_Ib & L\_Ib & Average & \multicolumn{1}{c}{R\_II+III+Va} & \multicolumn{1}{c}{L\_II+III+Va} & \multicolumn{1}{c}{R\_IV+Vb+Vc} & \multicolumn{1}{c}{L\_IV+Vb+Vc} & \multicolumn{1}{c}{R\_Ib} & \multicolumn{1}{c}{L\_Ib} & Average \bigstrut\\
    \hline
    Space & 78.06±7.87 & \textbf{76.63±7.67*} & 69.02±11.83 & 69.86±9.79 & \textbf{35.10±11.02*} & \textbf{34.31±10.18*} & \textbf{60.50±8.29*} & 61.66±10.48 & \textbf{57.62±12.17*} & 46.39±12.00 & \textbf{46.87±10.50} & \textbf{30.37±8.54*} & \textbf{30.95±8.59*} & \textbf{45.64±8.26*} \\
    17trophy & \textbf{78.18±7.90*} & 76.48±7.53 & \textbf{70.47±11.75} & 69.12±10.90 & 32.67±9.16 & 31.75±8.92 & 59.78±8.06 & \textbf{62.23±11.08*} & 56.79±11.92 & \textbf{48.15±13.63} & 44.67±12.60 & 27.92±7.28 & 27.83±7.29 & 44.60±8.40 \\
    Mules and Horses & 77.67±7.84 & 76.05±7.79 & 69.49±11.81 & 69.99±10.10 & 33.60±9.99 & 32.64±9.14 & 59.91±8.10 & 61.48±10.72 & 55.97±12.43 & 47.12±12.27 & 46.29±11.37 & 28.50±7.70 & 28.61±7.29 & 44.66±8.06 \\
    FIT-Monash & 77.70±8.06 & 75.88±7.41 & 69.54±11.39 & 69.84±10.13 & 32.98±9.12 & 32.33±9.25 & 59.71±8.04 & 60.90±11.15 & 55.75±12.12 & 46.68±12.10 & 45.51±10.99 & 28.09±7.08 & 28.37±7.83 & 44.22±8.10 \\
    NPU\_SAIIP & 77.77±7.72 & 75.87±7.29 & 69.48±11.05 & 68.74±9.99 & 33.41±9.91 & 31.59±9.36 & 59.48±8.03 & 61.15±10.67 & 56.13±11.64 & 46.79±10.93 & 44.81±9.82 & 28.00±7.61 & 27.63±7.84 & 44.09±7.80 \\
    Winwin & 77.07±8.02 & 76.27±7.34 & 69.63±11.37 & 69.65±9.87 & 31.98±10.09 & 32.05±9.36 & 59.43±7.98 & 58.92±11.06 & 56.21±11.63 & 46.29±11.29 & 45.34±11.20 & 27.55±7.74 & 27.76±7.55 & 43.68±7.78 \\
    junqiangmler & 76.44±7.75 & 73.55±7.83 & 66.20±11.98 & 65.18±11.96 & 34.00±9.33 & 31.42±9.50 & 57.80±7.68 & 60.28±11.48 & 54.47±13.07 & 44.60±12.22 & 43.28±11.81 & 27.90±7.11 & 27.32±8.11 & 42.97±8.30 \\
    SJTU\_LAB426 & 77.68±7.68 & 75.01±6.96 & 55.08±29.77 & 53.44±28.50 & 32.10±9.31 & 32.35±9.53 & 54.28±11.73 & 60.90±10.20 & 54.52±11.10 & 36.27±20.74 & 32.85±19.40 & 27.67±7.14 & 29.06±8.72 & 40.21±8.72 \\
    VIP-LAB & 75.02±8.43 & 73.50±8.62 & 67.55±10.95 & 67.29±9.52 & 29.29±8.64 & 29.21±8.59 & 56.98±7.74 & 57.70±11.93 & 53.71±12.73 & 44.22±11.02 & 42.42±10.73 & 25.86±7.28 & 26.21±7.67 & 41.69±8.34 \\
    cemrg & 48.07±7.38 & 42.91±9.02 & 29.27±16.67 & 18.98±13.39 & 8.69±4.47 & 7.26±5.66 & 25.86±6.07 & 26.80±6.65 & 24.04±7.42 & 16.80±9.04 & 9.92±6.63 & 12.04±4.71 & 13.39±5.49 & 17.17±4.27 \\
    \hline
    Baseline & 73.44±13.28  &  71.53±12.57  &  69.53±11.48  &  \textbf{70.01±10.05}  &  30.8±13.32  &  28.45±13.11 & 57.29±9.98 & 57.66±13.44  &  52.90±13.98  &  46.39±12.20  &  46.11±11.11  &  25.6±9.97  &  24.81±10.57 & 42.24±9.14 \\
    \hline
    \hline
    \multirow{3}[6]{*}{Method} & \multicolumn{14}{c}{ncCT-only subset} \bigstrut\\
\cline{2-15}          & \multicolumn{7}{c|}{DSC (\%)}                 & \multicolumn{7}{c}{NSD (\%)} \bigstrut\\
\cline{2-15}          & R\_II+III+Va & L\_II+III+Va & R\_IV+Vb+Vc & L\_IV+Vb+Vc & R\_Ib & L\_Ib & Average & \multicolumn{1}{c}{R\_II+III+Va} & \multicolumn{1}{c}{L\_II+III+Va} & \multicolumn{1}{c}{R\_IV+Vb+Vc} & \multicolumn{1}{c}{L\_IV+Vb+Vc} & \multicolumn{1}{c}{R\_Ib} & \multicolumn{1}{c}{L\_Ib} & Average \bigstrut\\
    \hline
    Space & 74.09±14.18 & 73.12±14.30 & 64.12±16.13 & 65.17±15.70 & 29.85±8.52 & 32.18±8.39 & 56.42±10.46 & 55.94±10.35 & 54.78±9.78 & 43.75±11.21 & 45.56±12.18 & 28.62±5.46 & \textbf{30.40±6.29*} & 43.18±6.31 \\
    17trophy & 75.21±14.34 & \textbf{73.48±14.18} & \textbf{66.06±16.22} & \textbf{65.23±15.83} & 30.57±8.07 & 32.80±7.25 & \textbf{57.23±10.28} & 58.43±9.35 & 54.78±8.96 & \textbf{46.53±11.09} & 44.66±11.56 & 27.43±4.84 & 29.37±4.30 & \textbf{43.53±5.55*} \\
    Mules and Horses & 74.96±14.28 & \textbf{73.48±14.19*} & 65.03±16.68 & 65.20±16.34 & 30.82±7.89 & 33.41±7.57 & 57.15±10.39 & 57.86±8.92 & \textbf{54.80±8.75} & 45.11±11.06 & \textbf{45.60±11.51*} & 27.36±5.15 & 30.12±4.41 & 43.47±5.68 \\
    FIT-Monash & 75.05±14.33 & 73.18±14.00 & 65.39±15.83 & 65.13±15.84 & 29.66±8.70 & 31.80±8.21 & 56.70±10.39 & 58.26±8.83 & 54.33±7.85 & 45.40±9.96 & 45.29±11.06 & 26.75±5.16 & 28.58±4.54 & 43.10±5.37 \\
    NPU\_SAIIP & \textbf{75.31±14.42} & 72.64±14.20 & 65.90±15.67 & 64.89±15.59 & 29.37±8.20 & 31.01±8.86 & 56.52±10.48 & \textbf{58.60±8.58*} & 53.65±9.03 & 46.00±10.22 & 44.84±11.96 & 25.41±4.64 & 26.73±5.18 & 42.54±5.72 \\
    Winwin & 73.52±14.21 & 71.63±13.93 & 63.11±16.57 & 63.84±16.48 & 28.48±6.84 & 29.76±6.49 & 55.06±10.16 & 54.55±10.86 & 50.78±9.04 & 42.53±13.72 & 43.65±12.78 & 26.07±4.63 & 26.64±3.68 & 40.70±6.75 \\
    junqiangmler & 74.22±14.20 & 70.87±13.68 & 64.49±15.35 & 63.95±15.07 & \textbf{35.13±9.10*} & \textbf{33.62±7.62*} & 57.05±10.09 & 57.56±8.09 & 51.41±8.29 & 44.91±10.61 & 43.94±9.68 & \textbf{29.56±5.30*} & 28.32±5.45 & 42.62±5.02 \\
    SJTU\_LAB426 & 68.08±18.88 & 69.22±18.92 & 44.03±31.61 & 35.17±33.08 & 26.26±9.20 & 30.29±9.58 & 45.51±14.95 & 48.31±15.35 & 50.43±14.11 & 29.74±20.91 & 22.84±21.73 & 23.91±7.24 & 25.90±6.96 & 33.53±10.47 \\
    VIP-LAB & 70.92±14.76 & 69.70±14.69 & 64.44±15.43 & 64.12±14.79 & 27.23±8.44 & 29.03±8.55 & 54.24±10.14 & 53.23±11.65 & 50.44±10.24 & 44.52±11.94 & 43.54±10.90 & 26.47±5.71 & 27.33±6.24 & 40.92±6.58 \\
    cemrg & 34.06±11.47 & 41.29±11.28 & 36.97±17.47 & 16.84±12.53 & 2.19±2.11 & 12.67±7.53 & 24.00±7.43 & 17.47±4.42 & 22.60±6.11 & 20.22±10.24 & 7.04±6.20 & 7.95±4.84 & 19.32±5.57 & 15.77±3.95 \\
    \hline
    Baseline & 70.28±19.73  &  68.21±19.38  &  64.29±16.41  &  63.35±16.65  &  29.54±9.95  &  29.19±11.55 & 54.14±12.79 & 53.94±13.86  &  51.10±13.51  &  44.03±10.45  &  42.75±11.70 &  24.72±6.95  &  26.22±7.53 & 40.46±7.90 \\
    \hline
    \hline
    \end{tabular}
    }
  \label{tab:LNCTV_DSC_NSD}
\end{table*}

\section{Results}\label{sec:set5}
\subsection{Results of Task01}
The quantitative results of GTVp and GTVnd segmentation on both internal and external testing cohorts are summarized in Table~\ref{tab:GTV_DSC_NSD} and the final rankings of all participating teams for Task01 are listed in Table~\ref{tab:GTV_Rank}. Fig.~\ref{fig:boxplot_task01} illustrates the patient-level distributions of DSC and NSD values. Most submitted methods achieved promising results on the internal testing cohort but exhibited significant performance decline on the external dataset, highlighting the challenge of achieving robust cross-center generalization in NPC GTV segmentation.

\subsubsection{Results on the internal testing set}
For GTVp segmentation, the \textit{Baseline} achieved mean DSC and NSD values of 77.53\% and 33.04\%, respectively. Among all participating teams, the \textit{Mules and Horses} team obtained the highest mean DSC and NSD of 79.47\% and 36.56\%, exceeding the \textit{Baseline} by 1.94 and 3.52 percentage points, respectively. The \textit{Winwin} and \textit{cemrg} teams demonstrated comparable performance. As shown in Fig.~\ref{fig:boxplot_task01}(a), eight out of ten achieved DSC values concentrated around 0.80, reflecting robust and stable segmentation across cases. In contrast, the NSD scores shown in Fig.~\ref{fig:boxplot_task01}(c) were substantially lower than the corresponding DSC values, suggesting severe under- or over-segmentation at object boundaries. 

For GTVnd segmentation, \textit{Mules and Horses} achieved the highest mean DSC of 69.74\% and \textit{Space} attained the highest mean NSD of 64.90\%, outperforming the \textit{Baseline} (DSC: 66.38\%, NSD: 61.96\%) by margins of 3.36 and 2.94 percentage points, respectively. As illustrated in Fig.~\ref{fig:boxplot_task01}(b), the patient-level DSC distribution for GTVnd was lower and more dispersed than that for GTVp, with median values around 0.70, indicating increased anatomical variability and greater segmentation difficulty. The NSD scores in Fig.~\ref{fig:boxplot_task01}(d) exhibited trends similar to those of the DSC. 

Qualitative visualizations in Fig.~\ref{fig:task01_vis_internal} show that the top-ranked methods were generally able to accurately localize GTVp and larger lymph nodes. However, precise boundary delineation, particularly for small lymph nodes, remains challenging.

\begin{figure*}[t]
    \centering
    \includegraphics[width=\textwidth]{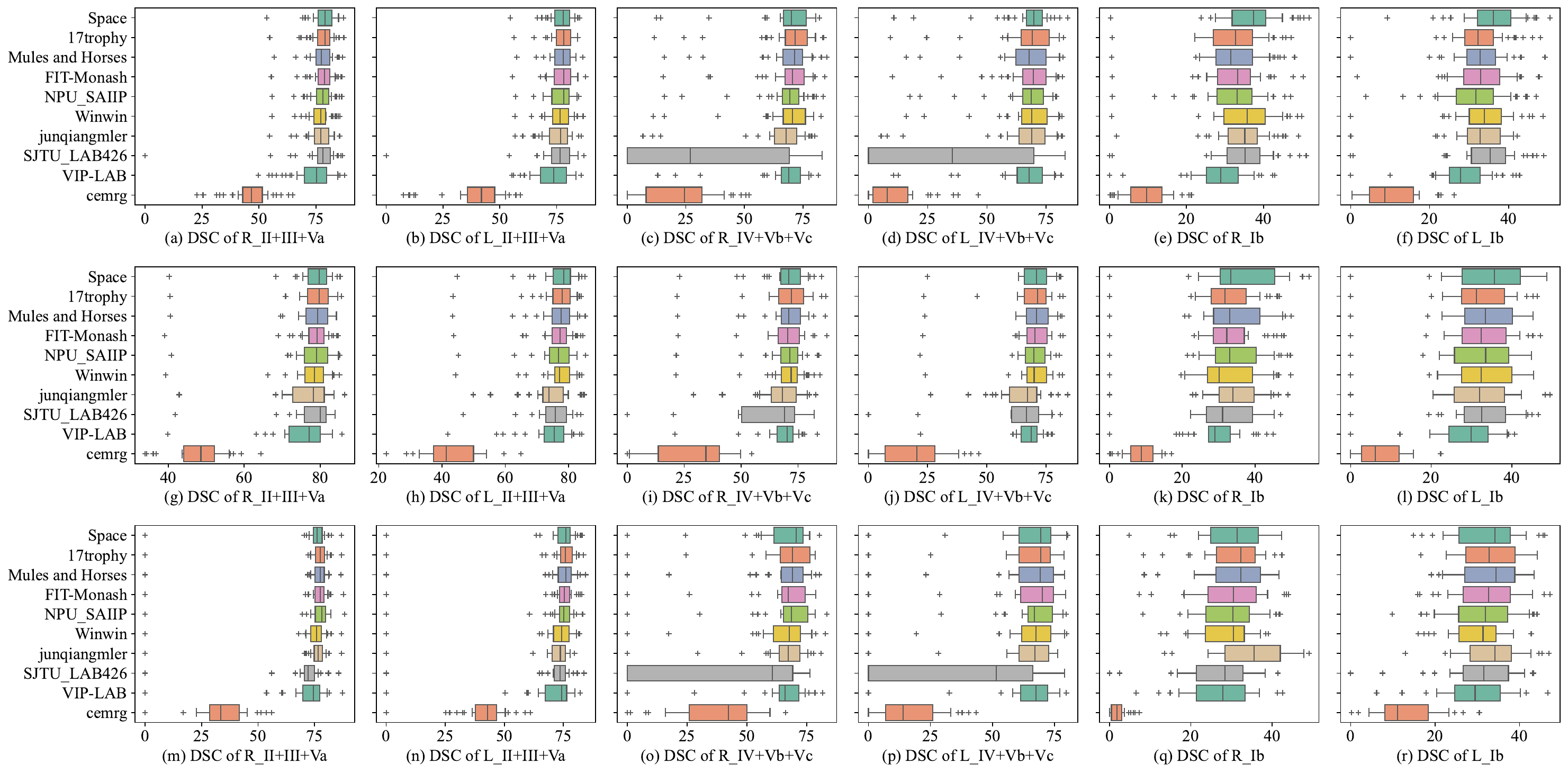}
    \caption{Boxplot of the patient-level average segmentation performance for LN CTVs in terms of DSC on (a)-(f) the paired CT subset, (g)-(l) the ceCT-only subset, and (m)-(r) the ncCT-only subset.}
    \label{fig:boxplot_task02_DSC}
\end{figure*}

\begin{figure*}[t]
    \centering
    \includegraphics[width=\textwidth]{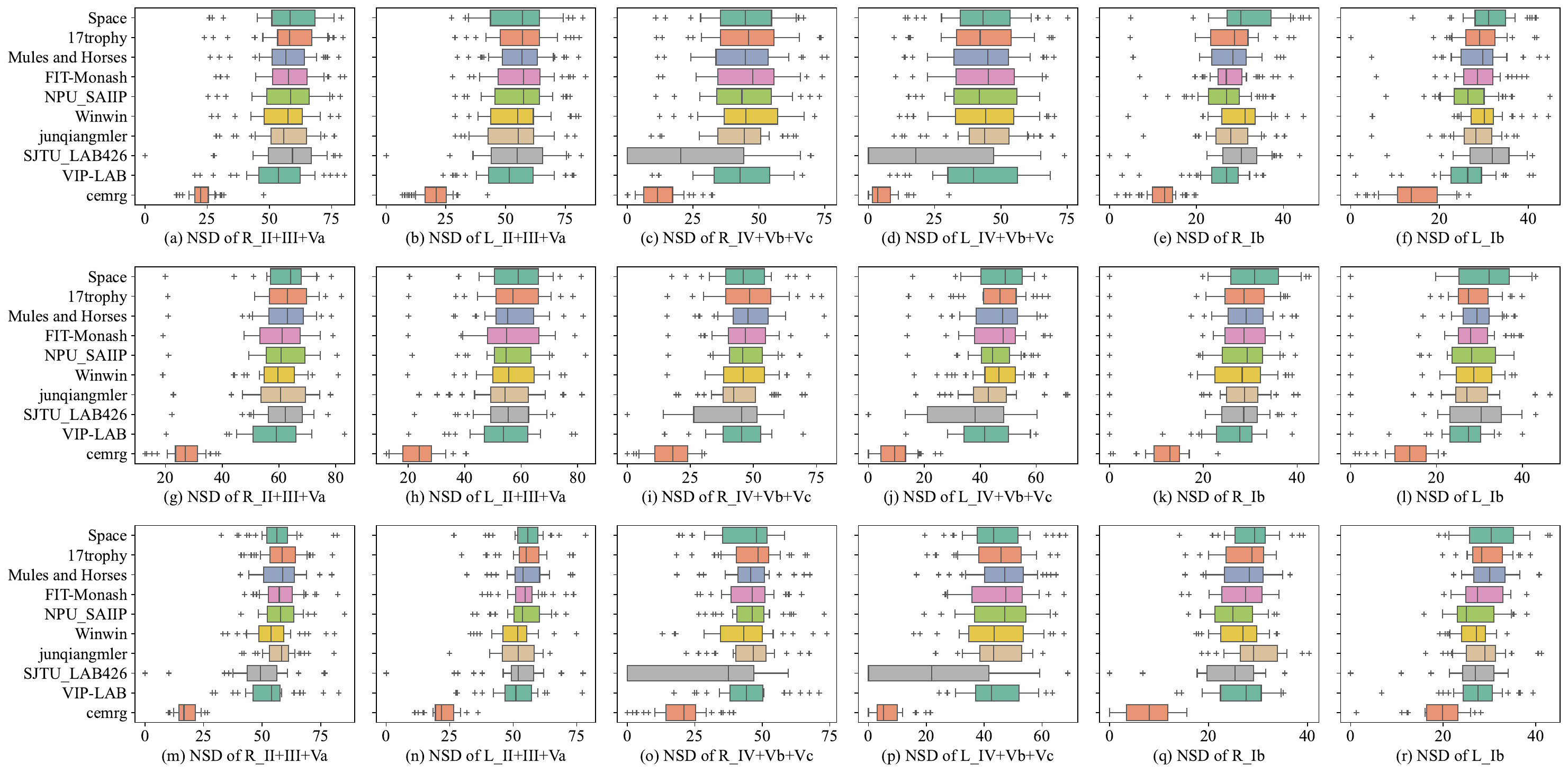}
    \caption{Boxplot of the patient-level average segmentation performance for LN CTVs in terms of NSD on (a)-(f) the paired CT subset, (g)-(l) the ceCT-only subset, and (m)-(r) the ncCT-only subset.}
    \label{fig:boxplot_task02_NSD}
\end{figure*}

\begin{figure*}[t]
    \centering
    \includegraphics[width=\textwidth]{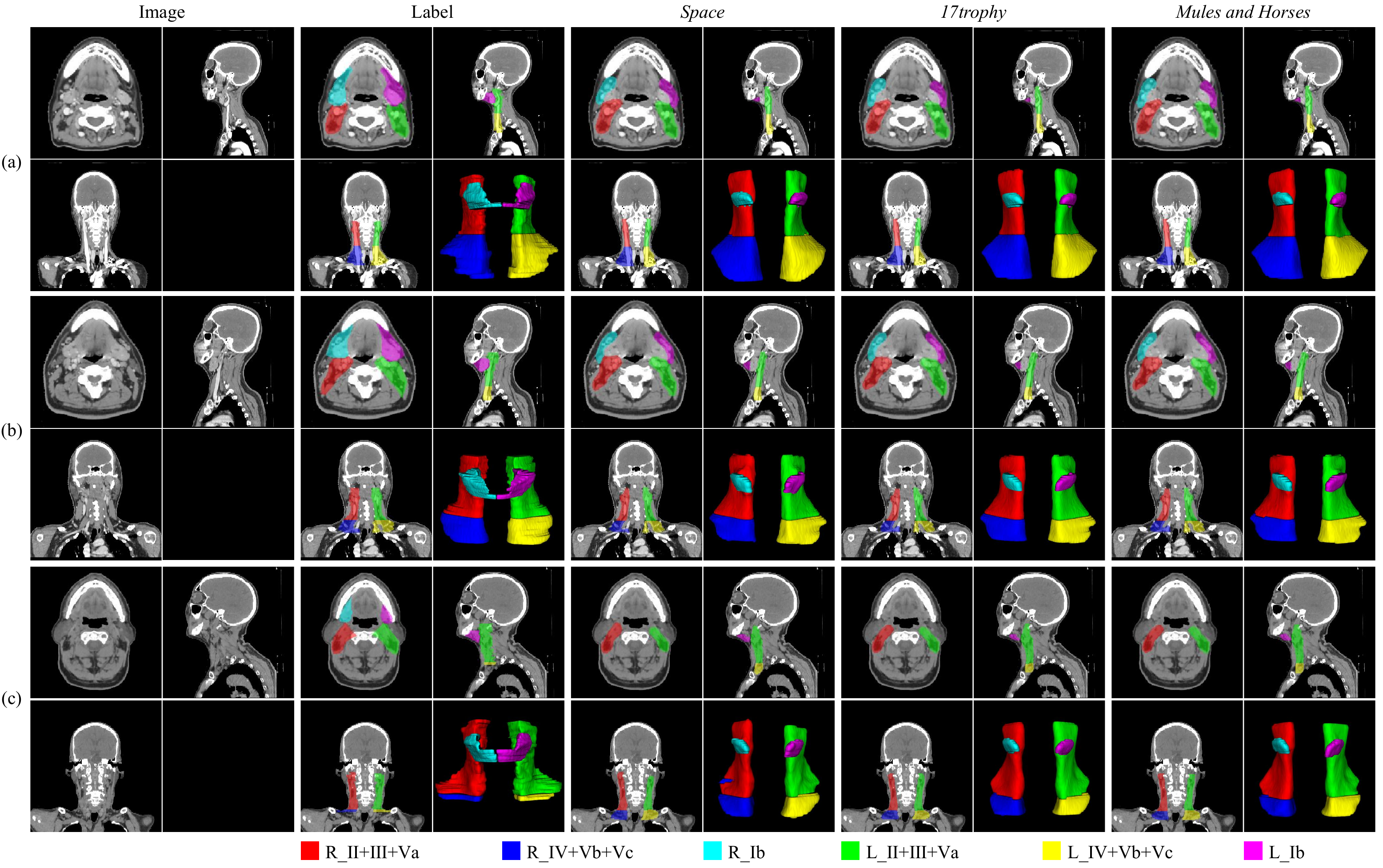}
    \caption{Qualitative GTV segmentation using the top three teams on (a) the paired CT subset, (b) the ceCT-only subset, and (c) the ncCT-only subset.}
    \label{fig:task02_vis}
\end{figure*}

\subsubsection{Generalization on the external testing set}
We further measured the segmentation performance of the participating methods on the external testing set to evaluate their generalizability. The results in Table 7 show that all methods suffered notable performance degradation on the external testing set compared with the internal testing set, reflecting considerable domain shifts across imaging centers. 

For GTVp segmentation, the \textit{Baseline} achieved mean DSC and NSD values of 48.19\% and 10.91\%, respectively, corresponding to decreases of 29.34 and 22.13 percentage points relative to the internal testing set. Six out of ten participating teams outperformed the \textit{Baseline} in both metrics. Among them, \textit{17trophy} achieved the best results, with mean DSC and NSD scores of 50.74\% and 14.54\%, respectively, outperforming the \textit{Baseline} by 2.55 and 3.63 percentage points. Fig.~\ref{fig:boxplot_task01}(e) and (g) indicate that both DSC and NSD values dropped substantially compared with the internal testing set, with median DSC values decreasing to approximately 0.50 and exhibiting noticeably wider interquartile ranges.

For GTVnd segmentation, the \textit{Baseline} exhibited reductions of approximately five and eight percentage points in mean DSC and NSD compared with the internal testing set, respectively, achieving mean DSC and NSD values of 61.19\% and 53.88\%. Among all participating teams, \textit{Space} achieved the highest mean DSC of 63.57\% and the \textit{FIT-Monash} team achieved the highest mean NSD of 58.14\%, outperforming the \textit{Baseline} by 2.38 and 4.26 percentage points, respectively. Unlike GTVp, the patient-level DSC scores of GTVnd shown in Fig.~\ref{fig:boxplot_task01}(f) remained relatively comparable to that of the internal cohort, with slightly lower NSD values observed in Fig.~\ref{fig:boxplot_task01}(h).

Qualitative results in Fig.~\ref{fig:task01_vis_external} demonstrate that the top three teams performed notable over-segmentation for both GTVp and GTVnd. These results indicate that automatic segmentation of GTVs remains highly challenging, especially when models are deployed across unseen external domains.

\subsection{Results of Task02}
The final rankings of all participating teams for Task02 are provided in Table~\ref{tab:LNCTV_Rank}, and the quantitative results for each LN CTV across three subsets are summarized in Table~\ref{tab:LNCTV_DSC_NSD}. Fig.~\ref{fig:boxplot_task02_DSC} and~\ref{fig:boxplot_task02_NSD} illustrate the distribution of patient-level DSC and NSD scores for each team. 

On the paired CT subset, most teams achieved stable and competitive results. The \textit{Baseline} achieved mean DSC and NSD scores of 54.89\% and 39.02\%, respectively. Eight out of ten teams surpassed the \textit{Baseline} in both metrics. The top-performing method, \textit{Space}, obtained mean DSC and NSD values of 60.23\% and 44.19\%, respectively, outperforming the \textit{Baseline} by 5.34 and 5.17 percentage points. As illustrated in Fig.~\ref{fig:boxplot_task02_DSC}(a)-(f) and Fig.~\ref{fig:boxplot_task02_NSD}(a)-(f), the top-five teams demonstrated consistent and stable segmentation performance on R\_II+III+Va and L\_II+III+Va, with median DSC values around 0.80 and median NSD values around 0.60. For the more anatomically complex levels IV, Vb, and Vc, the leading teams achieved median DSC values close to 0.70 and median NSD values around 0.45. Notably, \textit{STJU\_LAB426} failed to segment certain cases. For the levels Ib, the \textit{Space} team substantially outperformed all other methods, achieving noticeably higher median DSC and NSD scores.

On the ceCT-only subset, the overall segmentation performance was comparable or even better to that observed on the paired CT subset, suggesting that ceCT provides sufficient anatomical cues for reliable LN CTV delineation. The \textit{Baseline} achieved mean DSC and NSD scores of 57.29\% and 42.24\%, respectively, and seven out of ten teams exceeded these scores with stable results across LN levels. The top-performing team, \textit{Space}, achieved mean DSC and NSD values of 60.50\% and 45.64\%, outperforming the \textit{Baseline} by 3.21 and 3.42 percentage points, respectively. As depicted in Fig.~\ref{fig:boxplot_task02_DSC}(g)-(l), levels II, III, and Va remained the most stable LN levels, with top teams achieving median DSC values approaching 0.80. The distributions for levels IV, Vb and Vc were more compact than those on the paired CT subset, with slightly higher DSC and NSD scores (about 2 percentage points). For the levels Ib, the best-performing method, \textit{Space}, approached the performance obtained under the paired CT subset.

In contrast, on the ncCT-only subset, all methods experienced a noticeable decline in segmentation accuracy compared with the ceCT-only subset. The \textit{Baseline} obtained mean DSC and NSD scores of 54.14\% and 40.46\%, respectively, which were much lower than those on the ceCT-only subset. Eight out of ten teams still outperformed the \textit{Baseline}. The best-performing team, \textit{17trophy}, achieved mean DSC and NSD values of 57.23\% and 43.53\%, respectively, followed closely by \textit{Mules and Horses} and \textit{Space}. Fig.~\ref{fig:boxplot_task02_DSC}(m)-(r) and Fig.~\ref{fig:boxplot_task02_NSD}(m)-(r) indicates that the results on L\_IV+Vb+Vc were comparable to that on the paired CT subset, whereas the remaining LN CTVs showed more dispersed distributions with lower medians and wider interquartile ranges.

Overall, the submitted methods achieved comparable segmentation performance on the paired CT subset (highest DSC: 60.23\%, NSD: 44.19\%) and the ceCT-only subset (highest DSC: 60.50\%, NSD: 45.64\%). In contrast, a noticeable performance decline was observed on the ncCT-only subset (highest DSC: 57.23\%, NSD: 43.53\%). These findings indicate that ncCT alone provides limited soft-tissue contrast for consistent LN CTV delineation, resulting in reduced robustness and increased inter-team variability. Qualitative visualizations in Fig.~\ref{fig:task02_vis} indicate that the top three teams produced very similar predictions, where larger LN levels were generally well localized, but the boundaries remained challenging and often required major refinement, especially for the Ib levels.

\section{Discussion}

\subsection{Challenges and current status of GTV and LN CTV segmentation in NPC}
To address the challenges of anatomical variability, unclear boundaries, and missing modalities in NPC radiotherapy segmentation, participating teams explored several effective strategies for both GTV and LN CTV tasks. For instance, \textit{Space} incorporated spatial priors from surrounding OARs to ensure anatomically consistent predictions, highlighting the importance of contextual information. To enhance representation learning and target localization, \textit{Mules and Horses} finetuned a lymph node foundation model, while \textit{Winwin} employed adaptive region-specific loss functions. Data augmentation strategies also played a critical role. \textit{17trophy} employed a mixup strategy and \textit{FIT-Monash} relied on intensity-based augmentation to mitigate domain shifts across centers and modalities. In addition, \textit{Mules and Horses} and \textit{Winwin} removed horizontal flipping during LN CTV segmentation to respect the inherent bilateral symmetry of lymph node distributions. These approaches contributed to performance gains. As shown in Table~\ref{tab:benchmark_task01}, 8 out of the 10 participating teams outperformed the baseline on the internal testing set, and 7 of them surpassed the baseline on the external testing set. Table~\ref{tab:benchmark_task02} shows that 8, 7, and 8 teams outperformed the baseline on the three subsets of task 2, respectively. The results demonstrate that most of the participating methods successfully bring advances to the segmentation performance. Compared with the SegRap2023 benchmark, which achieved average DSC values of 78.56\% and 67.75\% for GTVp and GTVnd, SegRap2025 improved the average DSC to 79.47\% for GTVp and 69.74\% for GTVnd, respectively.

Despite these advancements, significant challenges persist, especially for cross-center generalization and the segmentation of complex substructures. Performance for GTV segmentation on the unseen testing domain exhibited significant drops, indicating that while strategies like intensity augmentation partially mitigated domain shifts, they were insufficient for robust generalization. Furthermore, segmentation performance varied substantially across different LN levels. While levels II, III, and Va achieved relatively high accuracy (mean DSC $>78\%$), difficult regions such as level Ib and levels IV, Vb, and Vc showed noticeably lower performance due to weaker anatomical boundaries and limited tissue contrast in ncCT scans.

Moreover, a clear gap persists between the current automated performance and the standards required for direct clinical application. Prior studies~\citep{liao2022automatic,luo2023deep,sjogreen2024landmark} suggest that DSC values exceeding 83\% for GTVp and 80\% for GTVnd, and 79\% for LN CTVs might be considered clinically applicable with few refinements. However, none of the submissions in this challenge met these criteria on either internal or external testing sets. These findings indicate that although top-performing solutions can accelerate the delineation process by producing automated segmentations within a few minutes, expert review and refinement remain indispensable to ensure clinical reliability and precision.

\subsection{Future trends}
To further improve segmentation performance and narrow the gap toward clinical applicability, several promising research directions can be explored for GTV and LN CTV segmentation. First, to better handle small or heterogeneous targets, future work could investigate multi-scale feature representations, coarse-to-fine cascaded architectures, and advanced loss functions (e.g., focal loss or reweighting loss) to alleviate class imbalance. Second, improving robustness to domain shifts remain a key priority. Integrating generalization-oriented strategies~\citep{yoon2024domain}, such as disentanglement learning, adaptive normalization, and self-supervised learning frameworks~\citep{wang2025volume}, may facilitate the learning of domain-invariant semantic representations. Third, addressing the missing-modality problem represents a promising direction, particularly through generative or diffusion-based approaches for synthesizing pseudo-contrast-enhanced CT from ncCT scans. Finally, incorporating multi-modal knowledge, such as introducing vision-language models and modeling spatial relationships among OARs, GTVs, and CTVs, may provide valuable contextual cues to guide more accurate and anatomically consistent segmentation.

\subsection{Limitations of this work}
Although SegRap2025 provides large-scale, multi-center, and multi-modality datasets for NPC GTV and LN CTV segmentation, some limitations still exist in the current work. First, the benchmark primarily targets pre-radiotherapy imaging, with limited inclusion of longitudinal scans (mid- and post-treatment) that are essential for evaluating model robustness during adaptive radiotherapy. Second, the dataset is restricted to imaging data and lacks complementary textual information, such as radiology reports or clinical records, which could provide valuable contextual cues. Third, the challenge focuses exclusively on contour delineation. Other clinically relevant tasks, such as dose prediction and treatment outcome modeling, have not yet been explored. In future work, we plan to expand the benchmark by incorporating additional data modalities, treatment stages, and a broader range of clinically relevant tasks to foster the development of comprehensive radiotherapy models.

\section{Conclusion}\label{sec:set7}
This work summarizes the outcomes of the SegRap2025 challenge, which provides a large-scale, multi-center benchmark for radiotherapy target segmentation in NPC. Building upon the original SegRap2023 dataset, SegRap2025 incorporates an unseen external testing set and adds a new LN CTV segmentation task that includes a mixture of paired and single-modality CT scans. Through systematic evaluation across institutions and imaging modalities, the challenge establishes a rigorous benchmark for assessing model robustness under real-world clinical variability. A total of ten teams submitted solutions for the GTV and LN CTV segmentation tasks, respectively. The results show that while the submitted methods achieve strong GTV performance on internal data, their accuracy degrades notably under domain shifts. For the LN CTV task, models benefited a lot from ceCT information, whereas segmentation based solely on ncCT proved considerably less reliable. These findings highlight that domain generalization and robustness to missing modalities remain unsolved challenges. Future research may explore domain adaptation methods and generative models to improve robustness across centers and imaging protocols. Incorporating richer anatomical context may further enhance the delineation of radiotherapy target volumes. Moving forward, we plan to expand the dataset to include additional patient sources, data modalities, anatomical sites, and radiotherapy tasks to support the development of more clinically reliable models.

\section{Acknowledgment}
This work was supported by National Natural Science Foundation of China (Grant 62271115, 82203197), Science and Technology Department of Sichuan Province (Grant 2022YFSY0055, 2026NSFSC0660), the Sichuan Provincial Cadre Health Research Project (Grant/award number: 2023-803) and the Radiation Oncology Key Laboratory of Sichuan Province Open Fund (2022ROKF04). {We would like to thank M.D. S.C. Zhang, M.D. W. Liao, M.D. Y. Zhao, M.D. C. Li and their team members for data collection, annotation, and checking}. We also would like to thank the support team of the Grand Challenge Platform and the MICCAI challenge organization team for their sincere help while hosting the challenge. We also would like to thank all participants for their active participation and working hard.

\bibliographystyle{model2-names.bst}\biboptions{authoryear}
\bibliography{refs}

\end{document}